\documentclass[journal]{IEEEtran}

\usepackage[sorting=none]{biblatex}
\bibliography{references.bib}

\usepackage{bbm}
\usepackage{color}
\usepackage{caption}
\usepackage[table,xcdraw]{xcolor}
\DeclareCaptionFormat{boldtable}{\textbf{#1} #2 #3}
\usepackage{pifont}

\usepackage{graphicx}
\usepackage{bbm}
\usepackage{eufrak}
\usepackage{yfonts}
\usepackage{wrapfig}
\usepackage{enumitem}

\usepackage{xcolor,soul,framed} 

\colorlet{shadecolor}{yellow}

\DeclareGraphicsExtensions{.pdf,.jpeg,.png}

\usepackage[cmex10]{amsmath}
\usepackage{array}
\usepackage{mdwmath}
\usepackage{mdwtab}
\usepackage{eqparbox}
\usepackage{url}
\usepackage[mathscr]{euscript}
\usepackage{subcaption}
\usepackage{dsfont}

\usepackage{mathtools}

\usepackage{threeparttable}
\newcommand\mybar{\kern1pt\rule[-\dp\strutbox]{1pt}{\baselineskip}\kern1pt}
\usepackage{pifont}
\DeclareMathOperator*{\argmin}{argmin}
\usepackage{dsfont}
\usepackage{footnote}
\usepackage{tabularx}
\usepackage{threeparttable}
\usepackage{amsmath,amssymb,amsfonts}
\usepackage{amsmath}
\usepackage{multicol}
\usepackage{tabto}

\usepackage{mathrsfs}
\usepackage{dblfloatfix}    

\usepackage[linesnumbered, ruled]{algorithm2e}
\usepackage{algpseudocode}

\usepackage{mathtools} 
\usepackage[hang,flushmargin]{footmisc}
\usepackage{lipsum}
\DeclareMathOperator*{\argmax}{argmax}
\usepackage{hyperref}
\hypersetup{
    colorlinks=true,
    linkcolor=blue,
    filecolor=magenta,      
    urlcolor=cyan,
    citecolor=red
}

\begin{document}
\bstctlcite{IEEEexample:BSTcontrol}
    \title{A Constrained Optimization approach for Ultrasound Shear Wave Speed Estimation with Time-Lateral Plane Cleaning}

      \author{Md. Jahin Alam, and Md. Kamrul Hasan
\thanks{\textbf{Md. Jahin Alam}, and \textbf{Md. Kamrul Hasan} are with the Department of Electrical and Electronic Engineering, Bangladesh University of Engineering and Technology (BUET), Bangladesh. E-mail: [\textbf{khasan@eee.buet.ac.bd} (M.K. Hasan: Corresponding Author), \textbf{jahinalam.eee.buet@gmail.com}] } 
}


\maketitle

\begin{abstract}

\textit {Objective:} Ultrasound shear wave elastography (SWE) is a noninvasive way to measure stiffness of soft tissue for medical diagnosis. In SWE imaging, an acoustic radiation force (ARF) induces tissue displacement, which creates shear waves (SWs) that travel laterally through the medium. Finding the lateral arrival times of SWs at different tissue locations helps figure out the shear wave speed (SWS), which is directly linked to the stiffness of the medium.
Traditional SWS estimation techniques, however, are not noise resilient enough handling data filled with jitter noise and reflection artifacts. 
\textit{Approach:} This paper proposes new techniques to estimate SWS in both time and frequency domains. These new methods optimize a loss function that is based on the lateral signal shift parameter between known locations and is constrained by neighborhood displacement group shift determined from the time-lateral plane-denoised SW propagation data. The proposed constrained optimization is formed by coupling neighboring particles' losses with a Gaussian kernel giving an optimum arrival time for the center particle by enforcing stiffness homogeneity in a small neighborhood to enable inherent noise resilience. The explicit denoising scheme involves isolating the transitioning SW profile in each time-lateral plane, creating a parameterized mask. Additionally, lateral interpolation is performed to enhance reconstruction resolution and obtain increased displacement groups to enhance the reliability of the optimization. 
\textit {Main Results:} The proposed noise robust scheme is tested on a simulation (US-SWS-Digital-Phantoms) and three experimental datasets: (i) Mayo Clinic CIRS 049 model phantom, (ii) RSNA-QIBA-US-SWS, (iii) Private data. The performance of the constrained optimization is compared with three time-of-flight (ToF) and two frequency-domain methods. The evaluated tests have produced visually and quantitatively ($\mathrm{CNR}$, $\mathrm{PSNR}$, mean inclusion, and background SWS) superior and noise-robust reconstructions compared to state-of-the-art methodologies. 
\textit {Significance:} Due to its high contrast and minimal error SWS map formation, the proposed technique can find its application in
tissue health inspection and cancer diagnosis.

\end{abstract}


%
\IEEEpeerreviewmaketitle


\section{Introduction} \label{Intro_section}
\IEEEPARstart{U}{ltrasound} elastography is an acknowledged method for $in-vivo$ estimation of tissue pathology. The main two categories of elastography that have gained traction in recent times are strain-imaging \cite{bamber1999ultrasound} and shear-wave imaging \cite{sarvazyan1998shear}. Shear-wave elastography (SWE) is particularly useful since it does not require external operator-dependent compression, is reproducible, and can provide the stiffness of soft tissue both qualitatively and quantitatively in a non-invasive manner. Recent studies have demonstrated that stiffness is a very crucial mechanical attribute \cite{frauscher2005prostate, lyshchik2005thyroid, thomas2007real, thomas2010significant, shiina2015wfumb} in distinguishing healthy tissues from tumorous areas (statistically significant difference between benign $(1.73\;[SD, 0.8]\; ms^{-1})$ and malignant $(2.57\;[SD, 1.01]\;ms^{-1}) (P \ge 0.001)$ \cite{yu2011differentiation}. SWE paves the way for the measurement of this parameter. This method makes use of an acoustic radiation force (ARF) \cite{nightingale2011acoustic} to generate shear waves in an elastic medium with small-amplitude mechanical perturbations. The generated shear waves propagate perpendicular to the ultrasound beam axis, for remotely inducing transient tissue displacements, as first proposed by Sarvazyan et al. \cite{sarvazyan1998shear}. The propagation of the shear waves is tracked by using an imaging modality such as ultrasound in US-SWE or MRI in Magnetic Resonance Elastography (MRE). Since stiffness has a direct relationship with the shear wave speed, the stiffness distribution at each point of the intervening tissue can be inferred by measuring how fast the shear wave reaches different lateral positions. Initially employed to stage liver fibrosis or characterize breast tumors, SWE is now fully being explored for the quantification of stiffness of the heart and arteries, thyroid nodule assessment, gastrointestinal wall diagnostics, and prostate abnormality screening. It is also promising for assessing disease severity and treatment follow-up of various musculoskeletal tissues, including tendons, muscles, nerves, and ligaments \cite{taljanovic2017shear}.

\begin{figure*}[h]
    \centering
    \centerline{\includegraphics[width=0.98\textwidth]{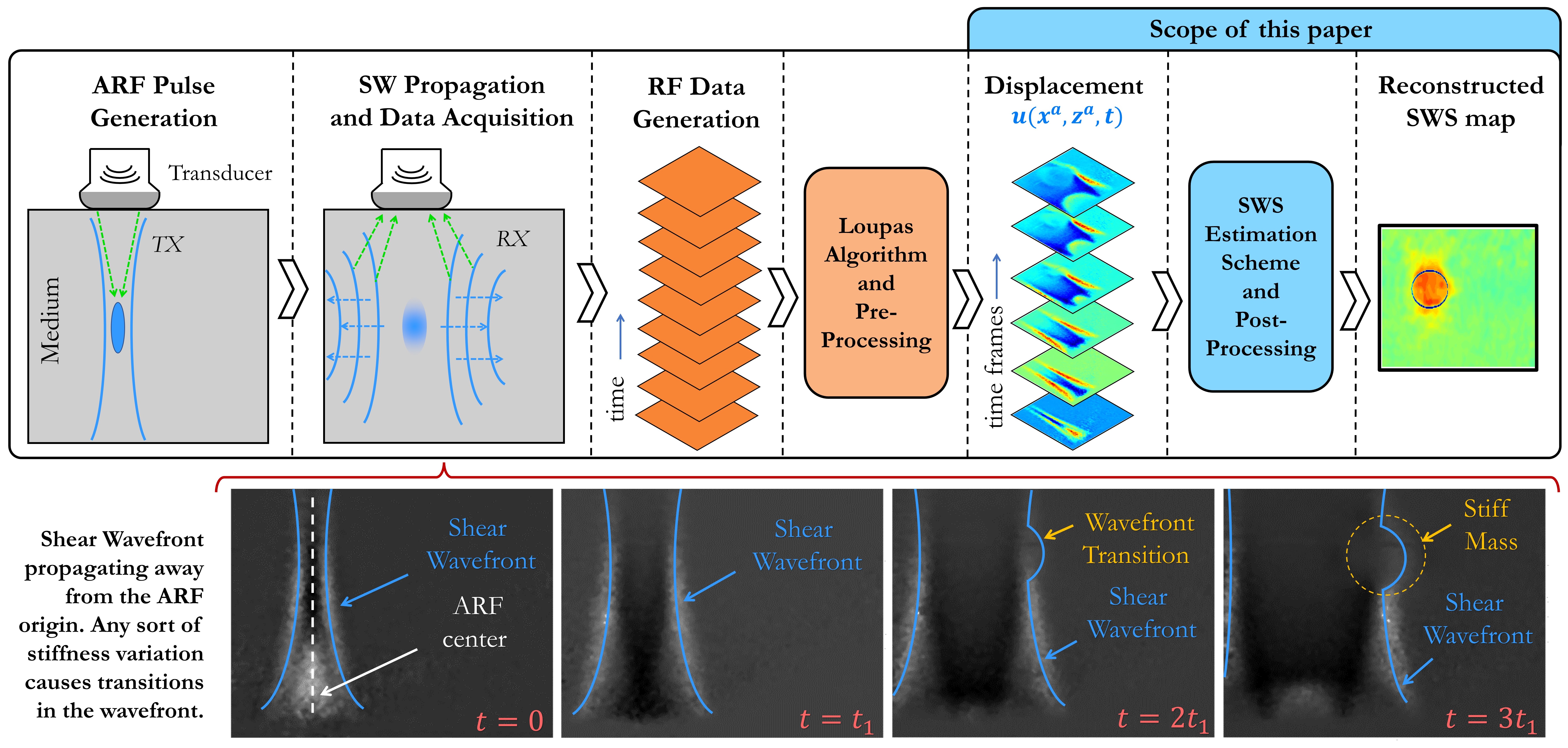}}
    \caption{{The sequential procedure of the Shear Wave Elastography (SWE) technique and visualization of the Shear Wavefront in a homogenous medium containing a stiff mass.}}
    \label{swe}
\end{figure*}

Figure  \ref{swe} depicts a typical Shear Wave Elastography reconstruction procedure. Using a transducer, an acoustic radiation force (ARF) is transmitted (TX) near a particular region of interest (ROI). The ARF beam (also known as push beam) is defined as follows:
\begin{equation}
    \mathrm{F} = \frac{2\alpha \mathcal{I}}{\mathcal{C}}
\end{equation}
Here, $\mathrm{F}$, $\alpha$, $\mathcal{I}$, and $\mathcal{C}$ are, respectively, the generated force, absorption coefficient, acoustic intensity, and speed of the shear wave. The effective force creates a displacement wavefront, depicted in figure  \ref{swe}, adjacent to the ARF center and gradually propagates throughout the medium longitudinally. This wave is known as the shear wave. Any spatial transition of the wavefront may indicate a probable tissue pathology change, e.g., lesion, cirrhosis, inside the region of interest (see $t=2t_1, 3t_1$ frames of figure  \ref{swe}). This is because the speed inside the transition is different from the other zones and this corresponds to a varied stiffness. With high-frequency plane wave imaging (PWI) the propagated shear waves are tracked (RX) and stored as RF data. 2D-Correlation-based Loupas algorithm \cite{loupas1995experimental, loupas1995axial} is implemented to detect the shear waves from the RF data and construct motion frames.  Using the wave transitions in the motion frames, the SWE reconstruction can be achieved with proper estimation techniques. However, the methods to create a viable distinction between normal tissues (liver: $<6kPa$, breast: $<45kPa$ \cite{goddi2012breast}\cite{mueller2010liver}) and possible abnormalities (fibrosis: $>8kPa$, tumor: $>45kPa$ \cite{goddi2012breast}\cite{mueller2010liver}) need to be precise. 

The preliminary investigations aimed to determine the shear wave speed (SWS) by utilizing the wave arrival period. These are commonly referred to as time-of-flight (ToF) techniques. The arrival period refers to the duration it takes for a shear wave to travel between two specified points. The time-to-peak (TTP) technique, frequently used, involves determining the peak time of displacement signals and performing a linear regression on the graph of peak time against distance. Slopes from this regression are calculated to obtain the SWS. Such an approach was successfully utilized in SWE \cite{sandrin2003transient, palmeri2008quantifying, fierbinteanu2009acoustic, chen2013assessment}; however, it is very susceptible to outlier peaks due to low signal-to-noise ratio, and artifacts. Methods such as RANSAC \cite{wang2010improving}, and Slope-Averaging \cite{ahmed2016shear}, which are known as best-fitting approaches, can effectively mitigate the impact of outlier peaks by analyzing the linear slopes. However, reflection can generate displacement peaks that occur before or after the expected time, which might decrease the accuracy of these approaches. To capture the complete signal profile, rather than simply the peak points, the most straightforward approach is to compute the cross-correlation between pairs of signals. Tanter et al. \cite{tanter2008quantitative} assume that the locations of the signal pairs are already known. The authors next determine the maximum correlation index to calculate the arrival time between these locations. Joyce McLaughlin and Daniel Renzi \cite{mclaughlin2006shear}, however, adopt a method where they designate one of the signals as a reference with a predetermined shape and shift it to maximize the cross-correlation. However, due to the prevalence of artifacts, the arrival time becomes inconsistent even in a homogeneous region \cite{rouze2012parameters}. As such, even precise cross-correlation can lead to jumps in estimation. Finally, the mentioned ToF methods do not use any prior knowledge regarding the shear wave propagation profile. Even in homogeneous tissue, these ToF reconstructions contain sporadic estimates of SWS for real-life settings, making them highly reliant on post-processing.


Transforming displacement fields into the frequency domain provides better selectivity for the signal delays and can be estimated from phase quantities. Chen et. al \cite{chen2004quantifying} have shown that there is a linear relationship between the phase delay ($\Delta \psi$) and propagation distance ($\Delta r$). The shear wave speed is then calculated, $\mathcal{C} = \omega \Delta r / \Delta \psi$ in an elastic material, where the medium vibrational frequency $\omega$ is to be known. However, this frequency is not only unknown in real scenarios but also contains harmonics. Kijanka and Matthew \cite{kijanka2018local} employed a window-based method called LPVI (Local Phase Velocity Estimation). They aimed to capture the phase sparsity of a center particle of the window and subsequently determine its shear wave velocity with respect to its surrounding area. Although their consideration of neighborhood effects had led to cleaner reconstructions, the outcomes hugely depend on the window size and frequency band selection; the process requires a lot of trial and error. Rosen and Jiang \cite{rosen2018fourier} proposed on using the full shape of the Fourier-domain waves and aligning them to increase the robustness of the SWS reconstruction (named FDSM: Fourier-Domain Shift Matching). They aim to shift and re-align a SW displacement signal to the ARF origin in the spatial or, temporal transformed domain by multiplying and iteratively changing a parameterized phasor. The iteration, where the phasor maximizes an alignment-function, produces the SWS. Such maximization approach is the first to be used in SWE and can work well for homogeneous mediums. But it will provide suboptimal results for heterogeneous regions having transitions is stiffness. Also, here a particular axial coordinate had been chosen during SWS estimation which does not take any neighborhood consideration into account.

Some loss functionality-oriented elastography schemes have been proposed \cite{ara2013phase}\cite{nahiyan2015hybrid}, but those are for strain ($\varepsilon$) image reconstruction. Ara et al. \cite{ara2013phase} determined an appropriate compression factor, $1/\alpha= 1/(1-\varepsilon)$, relating the pre- and post-compressed RF echo signals using the zero-lag phase between them. For strain continuity as well as reducing susceptibility to noise, an interrogative window was used over the phase values which decayed exponentially with distance. Nahiyan and Hasan \cite{nahiyan2015hybrid} have proposed a modified direct average spectral strain estimation technique where they calculated the maximum normalized cross-correlation (NCC) coefficient between the pre- and post-compressed RF echo signals. They determined the averaged NCC coefficient of a point with a 2D weighting function. The weighting function coupled the NCC coefficients from local tissue elements to the center point, with adjacent tissue elements being coupled strongly. The purpose of the neighboring NCC-weighted average approach was for local continuity and also minimization of decorrelation noise. Apart from FDSM, a metric or loss optimized estimator has not been introduced in Shear-Wave Elastography, especially one that incorporates neighborhood coupling. 


In this paper, novel constrained optimization-based noise-resilient techniques have been proposed to estimate tissue stiffness from SWS measurements in both time- and frequency-domain. A new loss function with constraints is optimized to determine the best lateral displacement signals' shift parameter of predefined locations. First, an objective function designed with the loss is taken to determine the best shift parameter of a particle of which the SWS is to be estimated. Next, this objective function is minimized subject to a constraint that the neighboring particles possess the same stiffness as the aforementioned particle. 
The coupling factors for the constraints is obtained from a spatial Gaussian kernel that basically exponentially weights a particle based on its distance from the centre. 
This idea essentially imposes stiffness homogeneity in a local region for enabling intrinsic noise robustness with the new loss function. Additionally, the time and frequency-based losses are combined to devise a dual-domain constraint optimization as a novel SWS estimation technique. The optimization of the displacement group shift is performed on SW propagation data which have been prior-denoised in the time-lateral plane of the motion data. The proposed denoising approach tracks the SW in each time-lateral plane utilizing piece-wise line fitting on the SW profile transitions. Then these lines are used to generate a parameterized mask to isolate the transitional SW profile. Furthermore, lateral interpolation is performed on the raw displacement data as a pre-processing step resulting in increased displacement groups for arrival-time measurements and improved reconstruction resolution. The simulation and CIRS phantom-based experimental results demonstrate the superiority of the proposed approach in comparison to the previous works published in the literature.

\section{Problem Formulation} \label{problem_form_section}
\subsection{Problem Identification}


The SWE imaging is rooted to the estimation of SWS by inducing a shear wave front in the elastic tissue medium with an acoustic radiation force (ARF). The shear wave enables the medium particles to be displaced perpendicularly ($z^a$) to the wave direction ($x^a$). Here, the superscript $a$ indicates the analog spatial domain. The imaging RF-data is transformed and pre-processed to obtain displacement fields, ${u}(x^a,y^a,z^a,t)$.  Under the assumptions \cite{palmeri2011acoustic}\cite{lai2009introduction} that biological tissues are elastic, isotropic, and possess a linear relationship between stress-strain, the following expression holds \cite{oliphant2001complex}\cite{bercoff2004supersonic}:
\begin{equation} \label{c_eq}
\mathcal{C}  = \sqrt{\frac{\mu_s}{\rho}} = \sqrt{\frac{E}{2\rho(1+\nu)}}
\end{equation}
Here:\\
$\rho$ = Density of medium ($kgm^{-3}$)\\
$\mu_s$ = Shear modulus ($Nm^{-2}$)\\
$E$ = Elasticity modulus; stiffness measure ($Pa$)\\
$\nu$ = Poisson's ratio\\
$\mathcal{C}$ = Shear wave propagation speed (SWS) ($ms^{-1}$)\\
$\nabla^{2}$ = Laplacian operator\\

Based on the local homogeneity of tissue structures, ${u}(x^a,y^a,z^a,t)$ can be converted into a 2D ($x^a, z^a$) signal due to the shear wave's directional property. Also, since the Poisson's ratio and density for soft tissue are close to $0.5$ and $1000 kg m^{-3}$ \cite{choi2005estimation}\cite{glozman2010method}, respectively, (\ref{c_eq}) can be simplified as

\begin{equation} \label{modulus_c_eq}
E \approx 3\rho \mathcal{C} ^2 = 3000 \mathcal{C} ^2
\end{equation}

Under the aforementioned assumptions, the tissue stiffness can be approximated (usually in the kPa range) from the propagating shear wave speed, $\mathcal{C} $. It should be noted that the stiffness calculation from (\ref{modulus_c_eq}) can vary significantly due to small over- or under-estimation of the quantity $\mathcal{C} $ due to their non-linear relationship. As a result, the SWS requires to be estimated as precisely as possible.  

\begin{figure}[h]
    \centering
    \centerline{\includegraphics[width=0.45\textwidth]{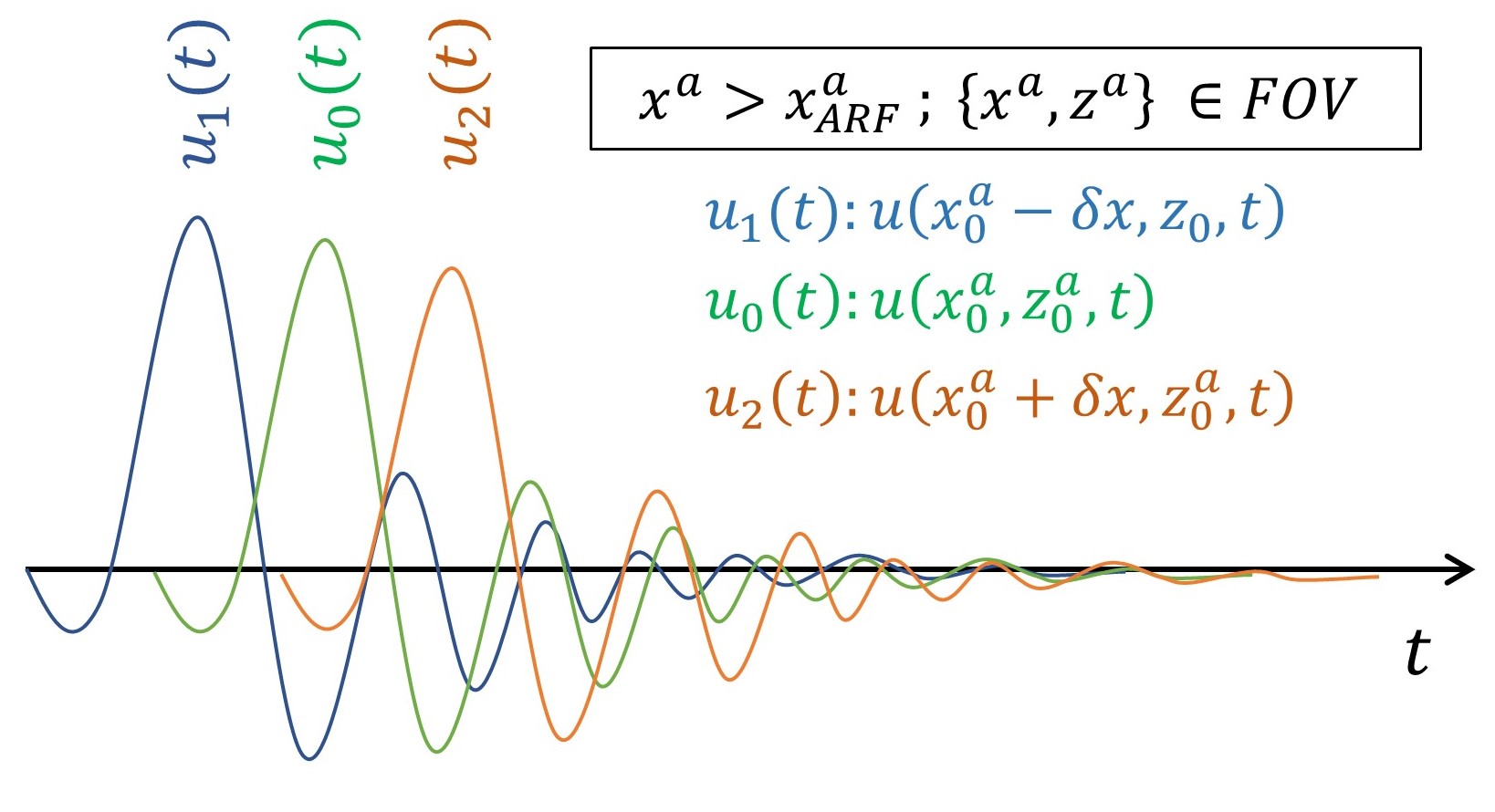}}
    \caption{{Displacements fields for three particles within FOV laterally separated by $\delta x$ from each other.}}
    \label{omega}
\end{figure}

\begin{figure*}[h]
    \centering
    \centerline{\includegraphics[width=0.98\textwidth]{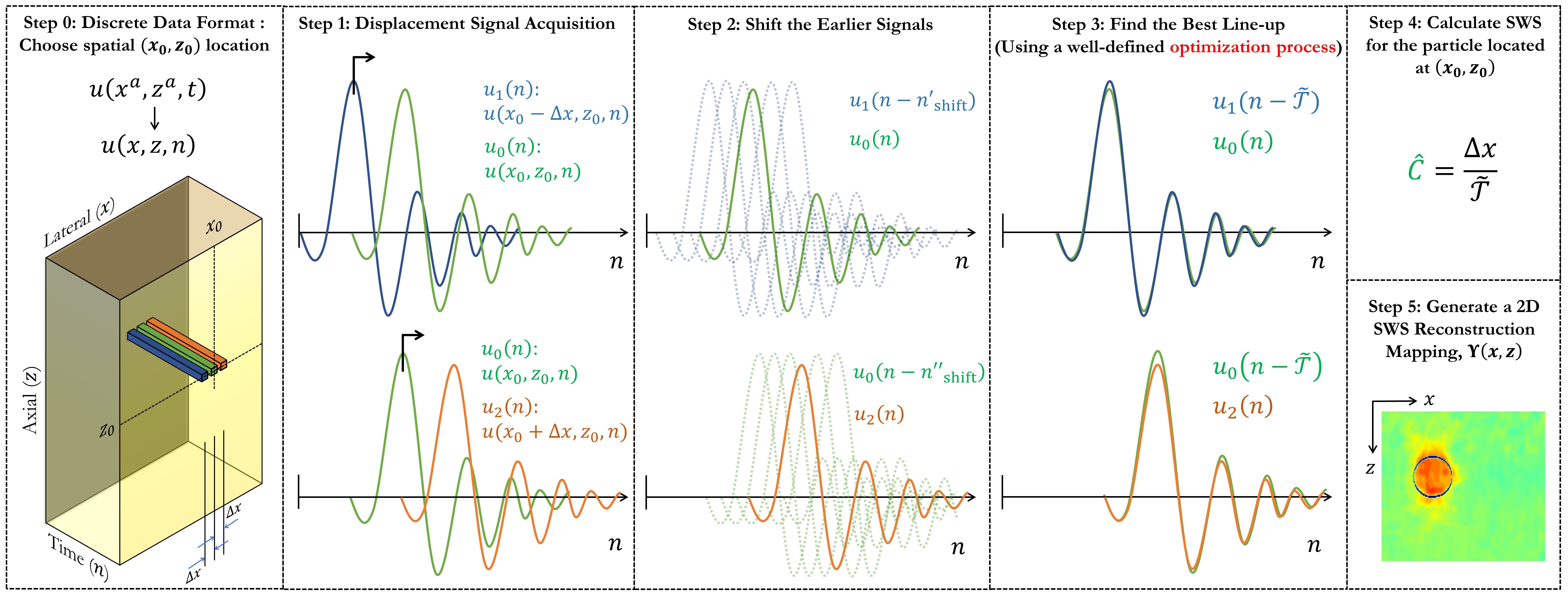}}
    \caption{Given a SWE displacement data structure, $u(x^a, z^a, t)$, the target is to determine a Shift-parameter $\tilde{\mathcal{T}}$ that best describes the arrival time between signal groups. Spatially separated by $\delta x$, the signal generated earlier can be shifted to line up with the delayed one. An optimization scheme is necessary to determine an optimum Shift-parameter $\tilde{\mathcal{T}}$ for all local signal groups.}
    \label{Problem}
\end{figure*}

The displacement signal $u(x^a,z^a,t)$ has a distinct structure, reaching a peak value when a laterally propagating shear wavefront reaches a particle and then gradually decays out. The speed at which the wavefront reaches a particle located at $(x_0^a,z_0^a)$ directly affects the timing of the formation of $u(x_0^a,z_0^a,t)$, with faster arrival resulting in earlier formation. In contrast, a particle located at $(x_0^a+\delta x,z_0^a)$ is faced with displacement with some time delay ($\delta x$ in typically $mm$). Considering a SW is propagating from $-x^a$ to $+x^a$ direction, figure  \ref{omega} displays example displacement fields $u_i(t), i=0,1,2$ with $u_0(t)$ as the reference for three particles on the same axial ($z_0^a$) point but separated laterally by $\delta x$. 

There is a direct association between the local SWS and time of occurrence for displacement $u(t)$. As a result, the delays of the three signals $u_0(t), u_1(t),$ and $u_2(t)$ with regard to one another are important in SWS estimation. In theory, assuming energy loss as the shear wave propagates, the following can be written:

\begin{equation} \label{omega_tau_01}
u_0(t) =  \alpha_{10}^a\;.u_1(t-\Gamma);\; \alpha_{10}^a<1.0
\end{equation}
\begin{equation} \label{omega_tau_20}
u_2(t) =  \alpha_{02}^a\;.u_0(t-\Gamma);\; \alpha_{02}^a<1.0
\end{equation}
Here, $\Gamma$ indicates temporal-delay (typically in $ms$) between the displacement signals which is the same as the SW arrival time between a distance of $\delta x$. The three signals $u_0(t)$, $u_1(t)$ and $u_2(t)$ for a particular particle and $\delta x$ are referred to as a `signal group' in this paper. If $u_1(t)$ and $u_0(t)$ are appropriately shifted to align with $u_0(t)$ and $u_2(t)$, respectively, then in theory the quantity $\Gamma$ can be obtained. The corresponding SWS for the centre particle can therefore be calculated using the following expression:

\begin{equation} \label{final_C}
\mathcal{C} = \frac{\delta x}{\Gamma} \; [ms^{-1}]
\end{equation}


\subsection{Practical Considerations} \label{Practical_consideration}

With a known temporal frequency $F_s\; [Hz]$ and spatial frequency $F_{sp}\; [pixel/mm]$, the analog terms can be translated into the following discrete formats:

\begin{equation}
(x^a, z^a, t) \rightarrow  (x, z, n), \; \; (x_0^a, z_0^a) \rightarrow  (x_0, z_0)
\end{equation}
\begin{equation} 
\Delta x = \delta x . F_{sp}
\end{equation}
\begin{equation} 
\mathcal{T} = \Gamma . F_{s}
\end{equation}
\begin{equation} \label{omega_tau_01_discr}
u_0(n) =  \alpha_{10}\;.u_1(n-\mathcal{T});\; \alpha_{10}<1.0
\end{equation}
\begin{equation} \label{omega_tau_20_discr}
u_2(n) =  \alpha_{02}\;.u_0(n-\mathcal{T});\; \alpha_{02}<1.0
\end{equation}
\begin{equation} \label{final_C_discr}
\mathcal{C} = \frac{\Delta x}{\mathcal{T}} . \frac{F_s}{F_{sp}} \; [ms^{-1}]
\end{equation}

\begin{figure*}[h]
    \centering
    \centerline{\includegraphics[width=0.85\textwidth]{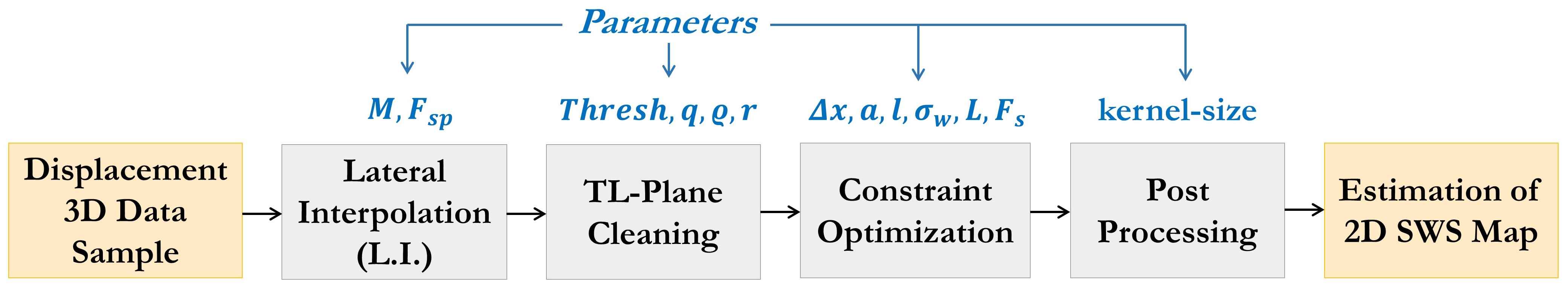}}
    \caption{{Graphical pipeline of the proposed methodology.}}
    \label{GA}
\end{figure*}

Additionally, for real-world shear wave elastography, there are two practical aspects to consider, both of which are due to noise and reflection artifacts:

\begin{itemize}
    \item (a) \textbf{Aspect-1}: In practice, $u_0(n), u_1(n), u_2(n)$ may not have the same shapes, especially the decayed out tail ends of the signals. 
    \item (b) \textbf{Aspect-2}: Secondly, although separated by the same lateral distance $\Delta x$, the delay periods between $u_0(n), u_1(n)$, and $u_2(n), u_0(n)$ may not be the same \cite{rouze2012parameters}. As such, unequal $\mathcal{T}$ quantities from (\ref{omega_tau_01_discr}) and (\ref{omega_tau_20_discr}) are plausible. 
\end{itemize}
The consequences of these aspects will be mitigated by the reduction of noise effects. Including estimates from nearest neighborhood homogeneous regions will lower the impact of noise. 


\subsection{Solution Approach and Target}

The approach of this paper is illustrated in figure  \ref{Problem} for solving the inverse problem at hand. The arrival time of a particle must be measured in order to estimate its SWS using 3D SWE data $u(x,z,n)$. To begin, the displacement field of the particle under consideration is acquired at any $(x_0, z_0)$, as well as two displacement fields from both lateral side particles at $(x_0-\Delta x, z_0)$ and $(x_0+\Delta x, z_0)$ (figure  \ref{Problem}, Step 1). Signal pairs formed by $\Delta x$ separated particles are obtained individually, and the earlier signals are iteratively adjusted to match up with their delayed counterparts (figure  \ref{Problem}, Step 2). The optimum lining-up from shifting will generate the best arrival time, $\tilde{\mathcal{T}}$ (figure  \ref{Problem}, Step 3). The target of this paper is defined in this particular step: find an optimization process to estimate the optimum $\tilde{\mathcal{T}}$ from signal groups that will best describe the arrival time for the center particle. For this, a constrained optimization approach is adopted. 

A constrained optimization problem involves minimizing an objective function, $f_0: \mathbb{R}^m \rightarrow \mathbb{R}_+$ (single or multi-variate, $v \in \mathbb{R}^m$), subject to a set of constraints $\mathcal{K}$ for the variables:

\begin{equation}
\tilde{v}= \argmin_v \; f_0(v), \textrm{ subject to } v \in \mathcal{K}
\end{equation}
Such a minimization approach is fitted to the SWE arrival time estimation task in question. The objective function is designed to produce the best line-up among a signal group of a particle, parameterized by the arrival time $\mathcal{T} \in \mathbb{R}^1$. The constraint $\mathcal{K}$ is formulated to address the noise influence. This is done by enabling the proximal neighborhood to have the same arrival time as the center particle (the particle dealt by $f_0$) and causing an ensemble effect. $\mathcal{K}$ can be represented with a constraint function ($h$) parameterized by the arrival time. Therefore, the constrained optimization problem  for the estimation task can be defined as follows:


\begin{equation} \label{constr_opt}
\begin{aligned}
\tilde{\mathcal{T}}= & \argmin_\mathcal{T}  \; f_0(\mathcal{T}); \; \mathcal{T} = 1, 2, ..., l_{sig}-1\\
\textrm{s.t.} \quad &  \argmin_\mathcal{S} \; h(\mathcal{S}) = \mathcal{T}; \; \mathcal{S} = 1, 2, ..., l_{sig}-1
\end{aligned}
\end{equation}
Here, $l_{sig}$ signifies the length of the discrete signals. Only one $\tilde{\mathcal{T}}\in \mathbb{R}^1$ will be obtained from the multiple neighboring signal groups. This will not be exactly equal to the true value, but it is meant to best predict the arrival time of the center particle while still producing a clean reconstruction, thus being the `optimum' shift parameter. $\tilde{\mathcal{T}}$ is used to estimate the SWS of the center particle, $\hat{\mathcal{C}}$ (figure  \ref{Problem}, Step 4). By repeating these four steps over all the spatial values of $(x_0,z_0)$ inside the field of view (FOV), the entire 2D-SWS reconstruction, $\Upsilon(x,z)$, can be generated (figure  \ref{Problem}, Step 5). The steps outlined above must be carried out while keeping the factors mentioned in sub-section \ref{Practical_consideration} in mind.

In short, given an experimental 3D SWE displacement data $u(x,z,n)$, a constrained optimization strategy can be designed to calculate the best-shift parameter $\tilde{\mathcal{T}}$ for each particle in the 2D-FOV and generate the 2D SWS mapping, $\Upsilon(x,z)$.



\section{Methodology}
In this section, research gaps in the existing classical and state-of-the-art methods have been presented as the motivation of this paper. Following that, the detailed planned scheme is shown, on which the SWS is estimated. The graphical pipeline of the proposed method is shown in figure  \ref{GA} with the parameters required for each step.

\subsection{Motivation} \label{motivation}
A crucial issue in classical ToF methods is that they are susceptible to inconsistent arrival times. For TTP \cite{sandrin2003transient, palmeri2008quantifying, fierbinteanu2009acoustic, chen2013assessment}, RANSAC \cite{wang2010improving} and Slope-Averaging \cite{ahmed2016shear} techniques, a linear regressive function is taken between the displacement peak-times ($t$) and the known lateral points ($x$). The inversed slope of the linear function provides the SWS of a particular point, $\mathcal{C}=(x_2-x_1)/(t_2-t_1)$. The arrival time of a single particle depends on the peak times of lateral particles. A slight distortion of these peaks (either temporally shifted or false-peak produced from noise) can change the slope and increasingly change the slope-inverse, resulting in over- or under-estimation. In the cross-correlation-based techniques, the entire shape of the displacement signals is considered. Now, assuming a pair of arbitrary signals, $v_1(n)$ and $v_2(n)$, is produced from the displacements of two laterally separated tissue locations, the arrival time can be calculated as
\begin{equation} \label{xcorr-eqn}
    \tau_{delay} = \argmax_{\mathcal{T} \in [-T,T]} \frac{1}{2T} \sum_{t=-T}^{T} v_1(n)v_2(n-\mathcal{T})
\end{equation}
where $\tau_{delay}$ is the maximum similarity delay index, $\mathcal{T}$. However, reflection artifacts can influence the similarity between two lateral signals and the maximum correlation calculated only between two signals may be found at an erroneous index. Here, a particle's SWS estimation depends only on $v_1(n)$ and $v_2(n)$ which, if faced with artifacts, will produce a sub-optimal arrival time.

There are various restrictions to consider for modern frequency domain-based estimate approaches. Firstly, LPVI \cite{kijanka2018local} employs a spatial-kernel-focused estimation approach. The technique transforms the time-axis to temporal frequency, selects a frequency band, and then converts the spatial-axes to spatial-frequency:

\begin{equation}
U^*(\Bar{x}, \Bar{z},F_0)= W_{x,z}(\Bar{x}, \Bar{z})\;.\;\mathscr{F}\left\{u(x,z,n) \right\} \big|_{F=F_0}
\end{equation}

\begin{equation}
\Tilde{U}(k_x, k_z,F_0)= \;\mathscr{F}_{2D}\left\{U^*(\Bar{x}, \Bar{z},F_0) \right\}
\end{equation}
Here, $\mathscr{F}$ and $\mathscr{F}_{2D}$ indicate $N$-point 1D and 2D discrete Fourier Transform (DFT), respectively. The dimension of the window $W_{x,z}(\Bar{x}, \Bar{z})$  and frequency band selection, i.e., $F_0$, are the two important parameters in LPVI. Higher frequency bands can be selected with smaller kernel sizes, but at the cost of the noise that exists in the high frequencies. Low $F_0$ requires higher kernel-size to capture sufficient spatial  transitions in $\mathscr{F}\left\{u(x,z,n) \right\} \big|_{F=F_0}$. Furthermore, the lower bound of the physical kernel-size needs to be at least $>4$~$mm$. This is not suitable for SWE data generated in low spatial resolution due to machine memory constraints.  In case of the FDSM \cite{rosen2018fourier} method, a SW $u(x) = g(x-\mathcal{C}t)$ is considered propagating away from the ARF being a shifted-version of $g(x)$. The delayed SW $g(x-\mathcal{C}t)$ is aligned to $g(x)$ in the Fourier-domain by zeroing out the shifting term $\mathcal{C}t$. Using the Time-Lateral plane, a dummy variable $\theta$ is used alongside $\mathcal{C}$ in the spatial frequency domain to do this. The Fourier Transform of the original shifted-signal and $\theta$ parameterized signals are, respectively, given by 

\begin{equation}
\begin{split}
\mathscr{F}\left\{u \left(x\right) \right\} = \mathscr{F}\left\{g \left(x-\mathcal{C}t \right) \right\} & = e^{-j2\pi k_x \mathcal{C} t} G(k_x) \\
& = U(k_x)
\end{split}
\end{equation}

\begin{equation} \label{FDSM}
\begin{split}
\mathscr{F}\left\{g \left(x-(\mathcal{C}-\theta)t \right) \right\} &= e^{j2\pi k_x (\theta-\mathcal{C}) t} G(k_x) \\
& = e^{j2\pi k_x \theta t} U(k_x)
\end{split}
\end{equation}
When $\theta=\mathcal{C}$, the shifted waveform will be aligned with the signal originating from $t=0$. The alignment phasor $e^{j2\pi k_x \theta t}$ can work very good if the unwrapped phase-line of $U(k_x)$ is linear. However, for real-world displacement signals, the phase lines are generally non-linear and especially noisy in high-frequency bands. The parameterized signal was integrated in \cite{rosen2018fourier} over all discrete time-instances for a better estimation. The best SWS has been calculated iteratively as  
\begin{equation} \label{FDSM_2}
    \hat{\mathcal{C}} = \argmax_{\theta} \sum_{m=1}^{p} \left| \sum_{n=-\frac{T}{2}}^{\frac{T}{2}} e^{j2\pi k_m  \theta n} U(k_m, n) \right|
\end{equation}
Here, $k_m$ and $p$ are the discrete spatial frequency index and total discrete spatial frequencies, respectively. For $\theta=\mathcal{C}$ the delayed signal is aligned and the magnitude in (\ref{FDSM_2}) becomes the maximum due to the $sinc$ function found by time-integration, as in (\ref{FDSM_3}): 
\begin{equation} \label{FDSM_3}
    \hat{\mathcal{C}} = \argmax_{\theta} \sum_{m=1}^{p} \left| \frac{sin \left( \pi k_m T (\theta-\mathcal{C}) \right) }{\pi k_m T (\theta-\mathcal{C})} G(k_m, n) \right|
\end{equation}
This maximization indicates a loss-functionality approach in total. But, if the signals have noisy oscillations, the alignment will be erroneous and the maxima of (\ref{FDSM_2}) cannot be identified. Furthermore, this approach relies on homogeneous mediums. Even if the locally homogeneous regions are used for estimation, the boundaries between inclusion and background can get blurred.


The slope-inverse techniques are based on linear regression of several signal peaks but do not take into account the full signal structure. And, the cross-correlation methods rely on only a pair of signals. Therefore, there is a scope to use the entire signal structure as well as signals from neighboring regions to reinforce the estimation, with the practical assumption of tissue homogeneity in physically small regions. Again, unlike the highly parameter-sensitive LPVI technique, the proposed method needs to be designed in such a way that the selection of the kernel-size does not need tuning, nor is the estimation sensitive to that kernel-size. The LPVI kernel is used spatially $(x,z)$ for a particular frequency ($F_0$) which is basically a 2D operation. But, the kernel can be used not only to acquire spatial $(x,z)$ signals but also to simultaneously optimize the signal groups' shift to a single value in time ($n$) or phase ($\omega=2\pi F$) domain, i.e., effectively a 3D operation based approach can be adopted. To mitigate the non-linearity problem of phase lines undisclosed in the FDSM method, a pre-possessing step solely for acquiring linear unwrapped phases as well as eliminating residual signal oscillations from the decaying displacement can be introduced. The shifting of the displacement signal groups will, therefore, produce identifiable extreme points from the optimization. Also, the kernel size is aimed to be set to $<2$~$mm$, which will cause low distortion to the inclusion-background boundary.  

The major contributions of this paper can be summarized as follows:

\begin{enumerate}
    \item The formulation of a time-domain Normalized Cross-Correlation (NCC) based loss function ensures the consideration of entire signal shapes; it is integrated into both the objective and constrained functions. 
    \item  The formulation of a phase-alignment-based Mean Squared Error (MSE) loss function enables increased selectivity from noisy phase components during optimized arrival time estimation; it is also integrated into both the objective and constrained functions.
    \item  The utilization of Gaussian-type weights with the objective and constrained functions provides spatial coupling while performing constrained shift-parameter optimization for local signal groups in unison.
    \item The development of a Time-Lateral Plane Cleaning (TL-Plane Cleaning) process performs piece-wise-line fitted SW profile cleaning to establish linearization of unwrapped phases as well as reduction of residual signals for better optimization.
    \item The qualitative and quantitative analysis confirms that the entire proposed methodology produces higher quality reconstruction compared to state-of-the-art ToF and Frequency-domain techniques from a performance point of view.  
\end{enumerate}

\subsection{Lateral Interpolation (L.I.)}

The transducers used for SWE have a fixed sampling rate (both temporal and spatial) during a data receiving cycle (RX, figure \ref{swe}) for image acquisition. Widman et al. \cite{widman2016shear} have experimentally shown while estimating arterial wall and plaque stiffness that a high Pulse Repetition Frequency ($PRF$) with low imaging quality is more important than low $PRF$ with better imaging quality. The $PRF$ is defined by

\begin{figure*}[t]
    \centering
    \centerline{\includegraphics[width=1\textwidth]{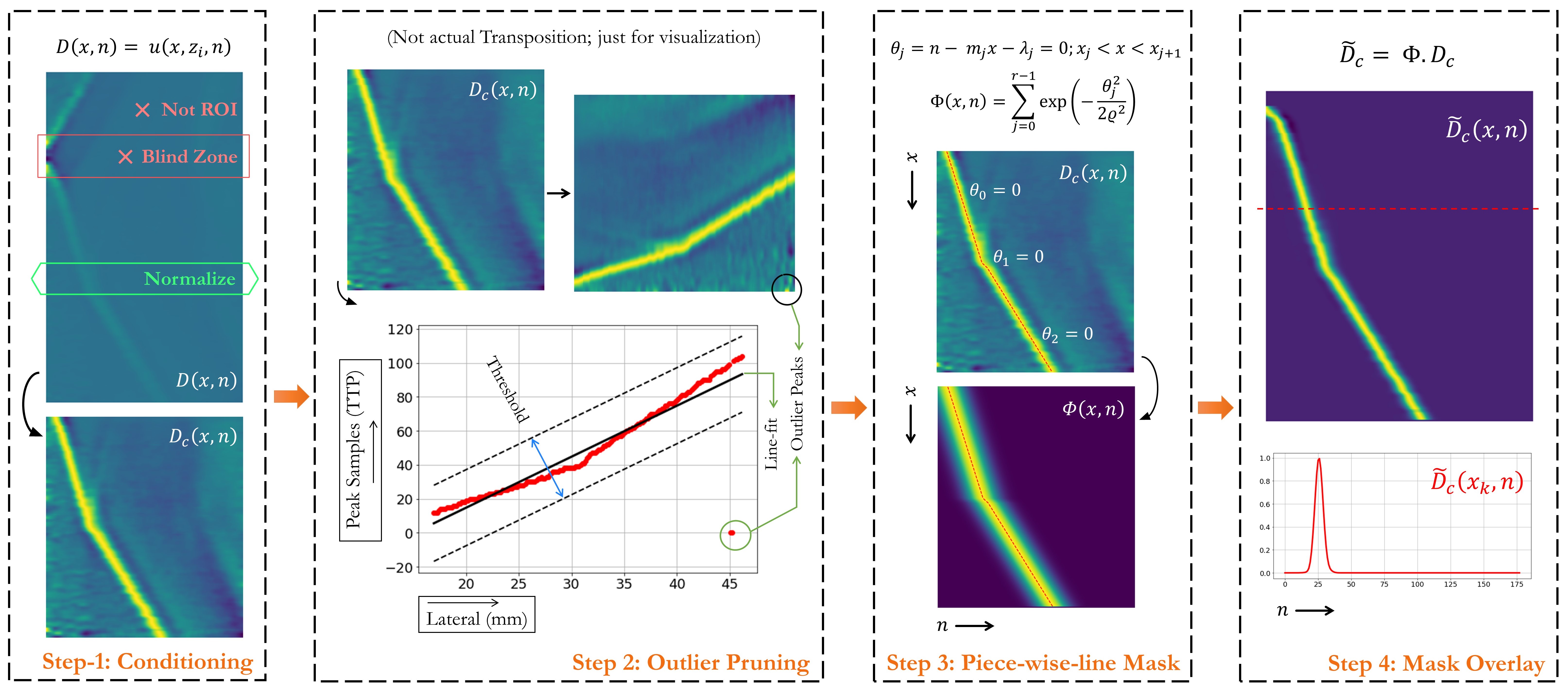}}
    \caption{{Time-Lateral Plane Cleaning: (1) omitting blind-zone and non-ROI regions. Then Normalization across the time-axis, (2) getting rid of outlier peaks with respect to a pre-defined threshold, (3) mask generation using Piece-wise line fit, (4) finally, overlaying the mask to attenuate everything else except the peaks.}}
    \label{TL_im_cleaning}
\end{figure*}

\begin{equation}
    PRF = \frac{1}{I_at_i} [Hz]
\end{equation}
Here, $I_a$ is the total imaging angles (higher $I_a$ improves imaging quality) and $t_i$ image acquisition time. The PRF controls the frame-rate of any ultrasound imaging and thereby, the temporal frequency. A high temporal frequency can be considered more important at the expense of spatial sampling frequency ($F_{sp}=\frac{1}{\mathrm{lateral\; resolution}}$ in $pixels/mm$) as it tracks the shear wave propagation through the tissue medium. Hardware constraineds may necessitate such a trade-off. This means that the physical distance between two laterally side-by-side particles (i.e., $x$ and $x+1$) in the pixel domain may vary among datasets imaged from different sources. The proposed optimization depends on the spatial neighborhood. Therefore, physically adjacent particles are utilitarian for optimizing the loss functions. In this paper, lateral points are interpolated using static displacements $(x, z=z_i, n=n_j)$ to acquire intermediate particles absent from the available data. Because shear waves move laterally, only the $x$-axis is taken into account. 

Suppose, a given SWE data has a particular dimension defined as $(lateral, axial, time)=(X, Z, T)$ where $ X, Z, T $ are integers. Static displacement amplitudes from $n=n_{j}$ are taken along the lateral axis from the data. Interpolation is performed following the process below:
\begin{equation} \label{lat_extract}
u_{ij}(x) = u(:,z_i,n_j)
\end{equation}
\begin{equation} \label{upsample_lat}
u^M_{ij} (x) = \mathscr{F}^{-1} \bigg\{\mathscr{F} \big\{u_{ij} \bigg(\frac{x}{M}\bigg) \sum_{i=\infty}^{-\infty} \delta(n-iM)\big\}H_M(e^{j\omega_x})\bigg\}
\end{equation}
\begin{equation} \label{upsample_lat_h}
H_M(e^{j\omega_x}) = \begin{cases}
      1 & \text{, $|\omega_x|< \pi/M $}\\
      0 & \text{, elsewhere}
    \end{cases} 
\end{equation}
The lateral interpolation order, $M$, is implemented with the spatial angular frequency $\omega_x$ to increase displacement values along the lateral axis.


%

\subsection{Time-Lateral Plane (TL) Cleaning} \label{TL_clean}
\begin{figure}[h]
    \centering
    \centerline{\includegraphics[width=0.49\textwidth]{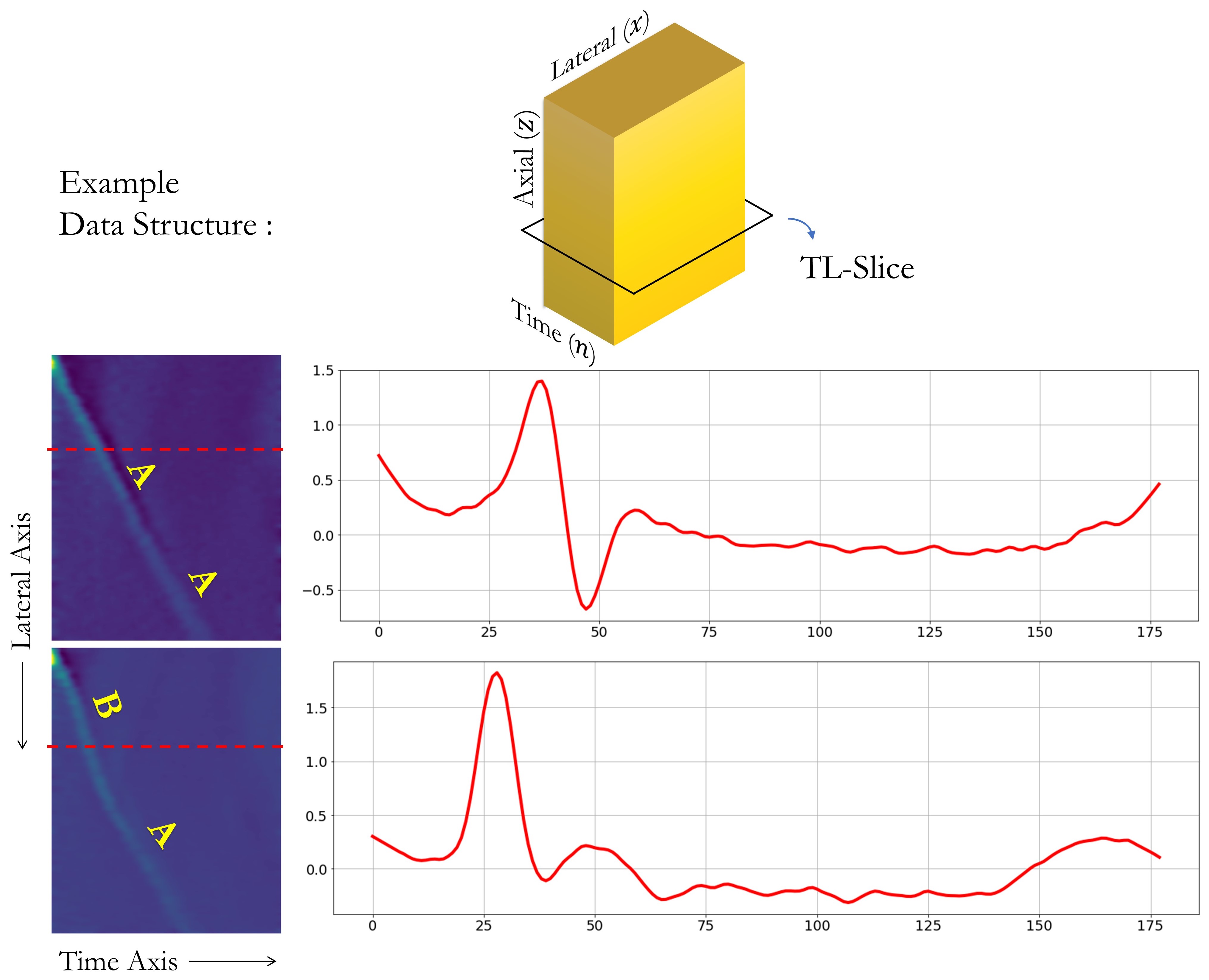}}
    \caption{{Two Time-Lateral Planes inspection selected from two particular axial points, taken as an example from a data structure. In the TL-planes, the dotted lines (\textcolor{red}{-\;-}) indicate a particular lateral point and the plots depict the displacement fields corresponding to the dotted line. The regions `A' and `B' denote different stiffness of the medium, `B' being higher.}}
    \label{TL_im0}
\end{figure}

\begin{figure*}[t]
    \centering
    \centerline{\includegraphics[width=0.99\textwidth]{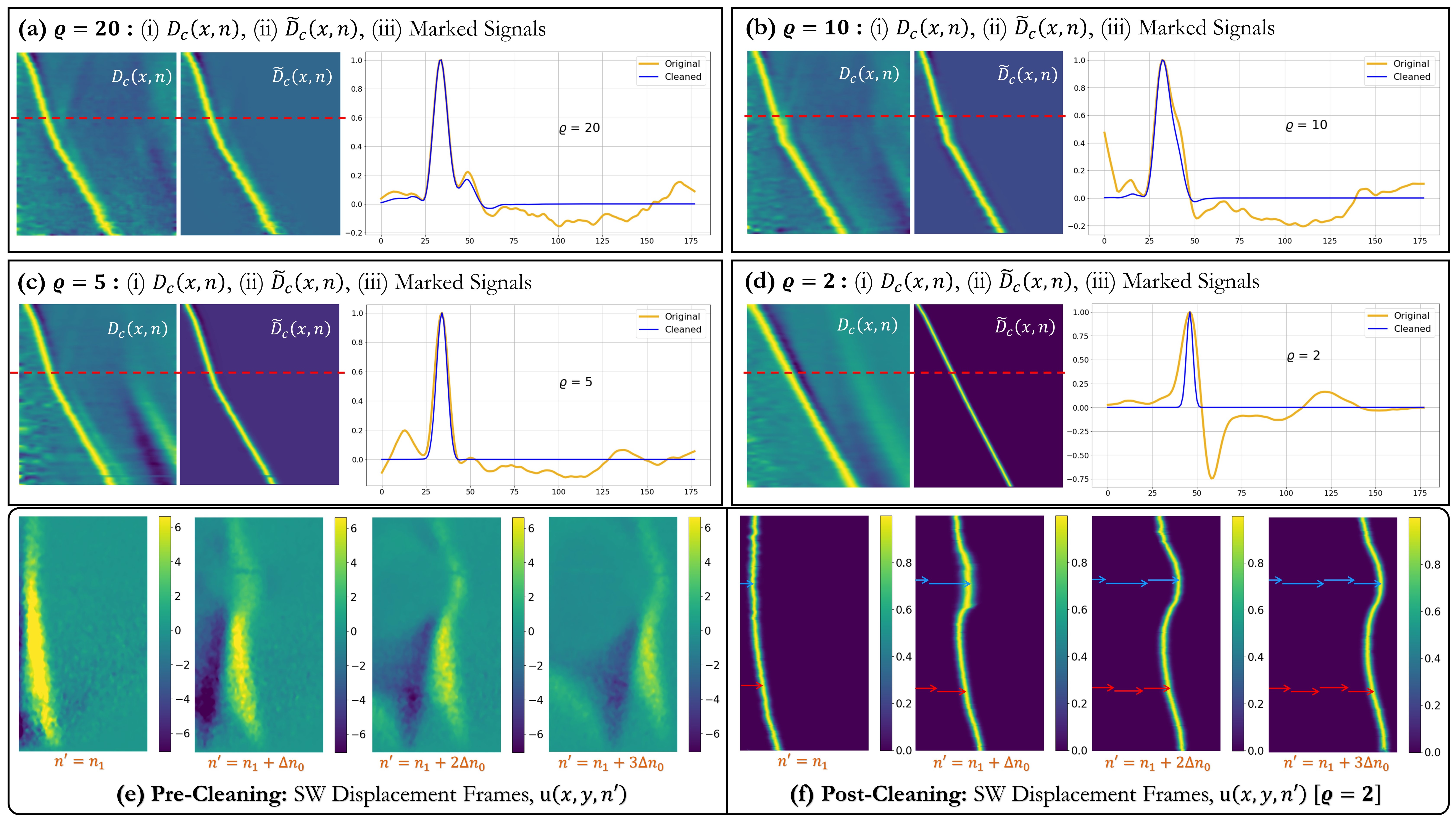}}
    \caption{{Instances (a-d) showing the effects of different quantities of $\varrho$ which control the cleaning extent in the TL-Plane Cleaning method. As the value $\varrho$ decreases $(20, 10, 5, 2)$, the Time-Lateral (TL) plane becomes sharper (contrast-wise), and the associating displacement signals become more selective with respect to the normalized peaks. Depictions in (e,f) shows an example of before- and after-effects of $\varrho=2$ TL-Cleaning of propagating SW frames (colorbar indicates displacement values). The size difference in the marked red and blue arrows indicate stiffness uniformity and variations, respectively.}}
    \label{TL_im_cleaning_demo}
\end{figure*}

For realistic data, the shear waves are influenced by jitter noise as well as reflection artifacts. Additionally, the particles will be faced with residual oscillations long after the shear wave has passed which carry no useful information. These residual oscillations pose contrast issues when the data is inspected as a video or an image. Additionally, they are detrimental to one of the constrained optimization approaches (explained in section \ref{Quantitative_Results}). Examples are shown in figure~\ref{TL_im0} procured from real SWE data. Two time-lateral slices are illustrated and amongst them, a lateral position is selected to examine the displacement signal. The signals possess residual values due to the particle displacements decaying out. In this paper, the tail-ended oscillations are eliminated with respect to the peaks through a newly proposed method, termed `Time-Lateral (TL) Plane Cleaning'.

There are some specific matters to take into account before tackling the cleaning process as stated below:

\begin{itemize}
    \item During the ARF push duration, the shear waves are very turbulent producing a blind zone where the displacement peaks cannot be tracked spatially. After that blind zone of around $(2-3) mm$ laterally, the shear wave signals begin to form their recognizable shapes.   
    \item The TL-slices indicate the direction of shear wave propagation. In a tissue medium, the stiffness can vary to various degrees enabling the shear wave to go slower (figure  \ref{TL_im0}, `A') or faster (figure  \ref{TL_im0}, `B'). This creates disjoint visual lines at the stiffness transitional areas.
    \item Due to noise, the visual line contributing pixels may become very much scattered around a mean trajectory, even when the line is supposed to be straight within a homogenous elasticity.      
\end{itemize}

The above-mentioned matters are each one-by-one handled in the TL-cleaning method. Each step of the proposed method is illustrated in figure \ref{TL_im_cleaning}. Firstly, a TL-slice, $D(x,n) = u(x,z_i,n)$, from a particular axial position $(z_i)$ is extracted. It is conditioned by defining the region of interest (ROI) of which the SWS is being investigated. This addresses the ARF issue mentioned above by excluding the blind zone and any additional area that is not under investigation. After the acquisition of the ROI, normalization with respect to the peaks is done along the time axis (horizontal axis in figure \ref{TL_im_cleaning}, Step-1). As a result, the intended ROI, $D_c(x,n)$, is obtained. However, there are still some outlier peaks that can be produced due to noise.

The next step deals with the noise issue. The time samples where displacement peaks have occurred (TTP) are inspected with respect to the lateral positions of the ROI slice. In the depicted figure  \ref{TL_im_cleaning} (Step-2), the horizontal axis indicates the lateral position $(mm)$. The vertical axis indicates the time samples when the displacement peaks occur; this is called the Time-to-peak (TTP). The ROI slice depicts a change of elasticity after $31$~$mm$. Some outlier peaks at the bottom of the plot are marked. Noise in SWE constitutes undesirable amplitudes to be added to the displacement signals and create these types of outlier peaks. The TL-cleaning will be performed using the peak locations and therefore the outliers are to be pruned from the ROI slice. For the pruning of outliers, a straight line is fitted to the entire TTP profile associated with a threshold ($T^{sh}$) range. If the squared distance of a peak from the fitted line is greater than $T^{sh}$, it is discarded. The threshold area must be tuned based on the extent of existing noise. After removing the outlying peaks, the next objective is to handle the discontinuous lines in the medium that correspond to stiffness transitions.


In Step-3, piece-wise lines are best fitted to each disjoint peak line. Some detected peaks still may not be within the expected trajectory due to the allowed threshold in Step-2. Therefore, all the signal quantities above a certain value $q=(0.8 - 0.95)$ are considered for the piece-wise line fitting. These lines (shown as \textcolor{red}{-\,-} in figure  \ref{TL_im_cleaning}, Step-3) are used as mapping functions to formulate 2D masks, $\Phi(x,n)$. The procedure mentioned above is mathematically presented below:

\begin{equation} 
m_j, \lambda_j \leftarrow Reg^1 \bigg[\left\{(x,n) : \arg D_c(x,n)>=q\right\}\bigg]
\end{equation}
\begin{equation} \label{line_piece_region}
\theta_j = n - m_jx- \lambda_j = 0 ; \;[x_j < x < x_{j+1}]
\end{equation}
\begin{equation} 
\Phi(x,n) = \sum_{j=0}^{r-1} exp \bigg(\frac{-\theta_j^2}{2\varrho^2}\bigg)
\end{equation}
\begin{equation} 
\Tilde{D}_c(x,n) = \Phi(x,n)\;.\;D_c(x,n) 
\end{equation}
Here, $m_j$ and $\lambda_j$ are the slope and intercept parameters, respectively, for piece-wise fitting a total of $r$ lines. Each $j$-th line (as such, $\theta_j$) is limited by the region defined in (\ref{line_piece_region}). Finally, the 2D masks are generated using the summation of each exponentially weighted $\theta_j$, the spread of which is controlled by the term $\varrho$. To obtain a clean TL-plane, the masks are overlaid on the ROI, $D_c(x,n)$, using element-wise multiplication. 

\begin{figure*}[t]
    \centering
    \centerline{\includegraphics[width=0.99\textwidth]{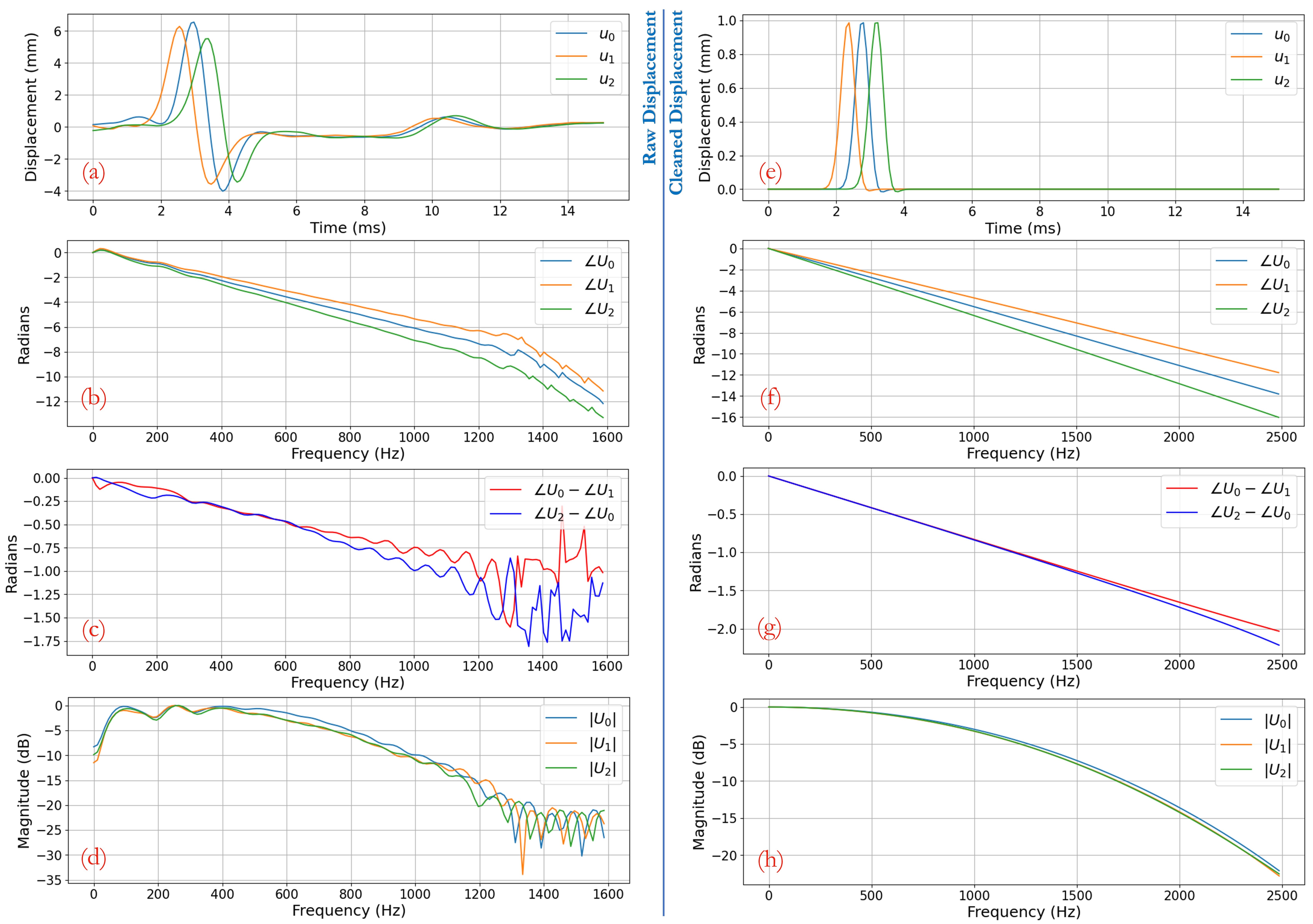}}
    \caption{{The raw and cleaned versions of the displacement signals $u_0(n), u_1(n), u_2(n)$ with their corresponding phase spectrums, phase differences, and normalized magnitude in dB.}}
    \label{phase_0}
\end{figure*}

The advantage of TL-Plane Cleaning is two-fold. Firstly, decreasing the value of $\varrho$ during cleaning eliminates residual quantities outside the peak indicating line to a greater extent as illustrated in figure  \ref{TL_im_cleaning_demo}(\textcolor{blue}{a-d}). This means that the tail-ends of the displacement fields are eliminated and replaced with near-zero quantities. The signals are also trimmed to some extent on both sides of the peaks. Secondly, in the frequency domain, the phase components become more and more linear as $\varrho$ decreases during TL-plane cleaning. This particular claim is illustrated in section \ref{phase_mse} through figure  \ref{phase_0} in detail. The contrast of the propagation frames of the principal shear wavefront is also greatly increased by TL-Plane Cleaning, as seen in figure  \ref{TL_im_cleaning_demo}(\textcolor{blue}{e,f}).

\subsection{Time-domain NCC Loss-based SWS Estimation} \label{time_ncc}

In the time domain, a loss function is to be formulated which addresses the two vital aspects mentioned in section \ref{Practical_consideration} as well as the motivation described in the section \ref{motivation}. The peak values of empirical signals can have any range based on time and location. All of them must be mapped to a known range with regard to each other, and the delay duration between the signals must be calculated directly. These two tasks can be effectively performed by the `Normalized Cross-Correlation' (NCC) function, defined as

\begin{equation} \label{ncc}
\mbox{NCC}\{a(n),b(n)\} =  \frac{\sum_{n=0}^{N-1} a(n).b(n)}{\sqrt{\sum_{n=0}^{N-1} a^2(n) . \sum_{n=0}^{N-1} b^2(n)} + \varepsilon} 
\end{equation}
Here, both $a(n)$ and $b(n)$ are discrete-time (DT) signals, sampled at a known temporal sampling frequency $F_s$. The constant $\varepsilon$ is a near-zero positive quantity in case one or both signals are zero everywhere. The reason for using NCC is that this metric does not measure the equality between two signals, rather it estimates the degree of similarity between them in the $0-1$ range (addresses aspect-1). The delay between the signals is provided by the sample index with the highest similarity. Considering the previously stated aspect-2, the delays of interest at a given coordinate $(x,z)$ are as follows:
\begin{equation} \label{ncc1}
\begin{split}
\tau_{10} & =  \argmax_\tau \bigg[\mbox{NCC}\big\{u_1(n-\tau),u_0(n)\big\}\bigg] \\ 
& = \argmax_\tau \bigg[\mbox{NCC}\big\{u(x-\Delta x,z,n-\tau),u(x,z,n)\big\}\bigg]
\end{split}
\end{equation}
\begin{equation} \label{ncc2}
\begin{split}
\tau_{02} &= \argmax_\tau \bigg[\mbox{NCC}\big\{u_0(n-\tau),u_2(n)\big\}\bigg]\\ 
& = \argmax_\tau \bigg[\mbox{NCC}\big\{u(x,z,n-\tau),u(x+\Delta x,z,n)\big\}\bigg]
\end{split}
\end{equation}
The obtained sample delays (both are integers and might be unequal in practice) provide an upper and lower bound among which the optimum sample shift resides. However, if $\tau_{10}= \tau_{02} \pm 1$, then the intermediate optimum shift value cannot be found unless each of the signals is interpolated. Hence, the upper and lower bound delay quantities $(\mathcal{T}_{10}, \mathcal{T}_{02})$ are calculated by initially up-sampling the displacement signals:

\begin{equation} \label{upsample}
u^L_k (n) = \mathscr{F}^{-1} \bigg\{\mathscr{F}\big\{u_k \bigg(\frac{n}{L}\bigg) \sum_{j=\infty}^{-\infty} \delta(n-jL)\big\}H_L(e^{j\omega})\bigg\}
\end{equation}
\begin{equation} \label{upsample_h}
H_L(e^{j\omega}) = \begin{cases}
      1 & \text{, $|\omega|< \pi/L $}\\
      0 & \text{, elsewhere}
    \end{cases} 
\end{equation}
\begin{equation} \label{T01_time}
\mathcal{T}_{10} = \argmax_\tau \bigg[NCC\big\{u^L_1(n-\tau),u^L_0(n)\big\}\bigg]
\end{equation}
\begin{equation} \label{T20_time}
\mathcal{T}_{02} = \argmax_\tau \bigg[NCC\big\{u^L_0(n-\tau),u^L_2(n)\big\}\bigg]
\end{equation}

The interpolated signals are produced using (\ref{upsample}) where $T$ is the finite signal length. Here, $\mathscr{F}$ and $\mathscr{F}^{-1}$ indicate $N$-point discrete Fourier transform (DFT) and inverse DFT, respectively, $L$ is the interpolation (or, upsample) order, and $H_L(e^{j\omega})$ is the anti-imaging filter corresponding to that order. Finally, $\delta(n-n_0)$ is the `Unit Impulse' or `Delta Function', having a unit value at $n=n_0$ and zero sample values everywhere else.

At this stage, a loss function is defined in the time-domain for SWS estimation. This loss assumes that the propagated shear wave moving from $(x-\Delta x, z)$ to $(x,z)$ has the same speed as that of from $(x, z)$ to $(x+\Delta x,z)$. Such an assumption is necessary for obtaining a single shift-parameter from practical data. The loss is formulated by negating the NCC terms and converting it into a minimization problem, as shown below (the subscript $TD$ indicates the time-domain approach): 
\begin{equation} \label{LTD}
\begin{split}
\mathscr{L}_{TD}(x,z|\Delta x, \mathcal{T})  = \; & 2  -\mbox{NCC}\{u_1^L(n-\mathcal{T}),u_0^L(n)\}\\
& -\mbox{NCC}\{u_0^L(n-\mathcal{T}),u_2^L(n)\}
\end{split}
\end{equation}
The shift-parameter $\mathcal{T}$ provides an overall goodness of fit between the \{shifted $u_1^L(n)$, un-shifted $u_0^L(n)$\} as well as \{shifted $u_0^L(n)$, un-shifted $u_2^L(n)$\}. Since the loss depicted in (\ref{LTD}) represents the particle of which the SWS is being estimated, it is regarded as the objective function, $f_0(\mathcal{T})$ {(from Eq. \ref{constr_opt})}. However, this particle under observation is a tangible part of the tissue medium. It should be homogeneous to its immediate neighborhood in real-life scenarios. To emphasize, a physical dimension of roughly $1-2$~$mm$ can be regarded as a region of homogeneity and thus the particles in this region can be used in the loss function to improve the robustness of the proposed SWS estimation method.


\begin{figure}[t]
    \centering
    \centerline{\includegraphics[width=0.48\textwidth]{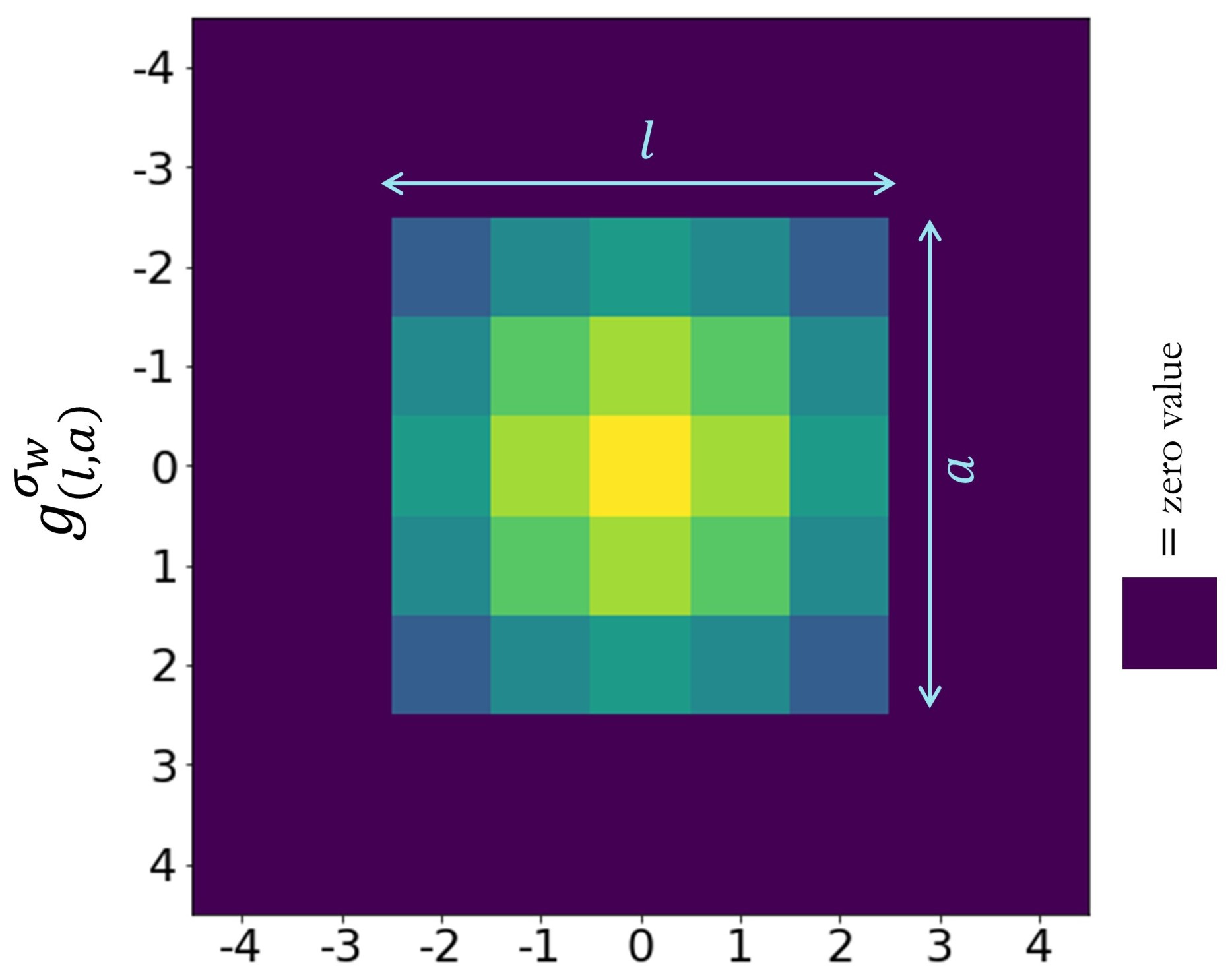}}
    \caption{{2D Gaussian function-based kernel which is to be integrated into the optimization. For visualization, $l=a=5$ and $\sigma_w=0.5$ are used. }}
    \label{gaussian_win}
\end{figure}

The constraint function, $h(\mathcal{T})$, is consequently set to the cumulative loss functional value produced from the neighborhood's signal shift. A preset $l\times a$ shaped (both $l, a$ are odd numbers) neighborhood is considered for this case. With spatial coupling factors $\mu_{k,i}$, losses from each particle are multiplied. The loss $\mathscr{L}_{TD}$ is notationally simplified with $\mathcal{J}_{TD}^{x,z}$ to depict signals from the local region. As a result, the optimization task is described as

\begin{equation} \label{constr_opt_LTD}
\begin{aligned}
\mathcal{T}_{TD} = & \argmin_{\mathcal{T}} \;  \mu_{0,0} \mathcal{J}_{TD}^{x,z}(0,0, \mathcal{T})\\
\textrm{s.t.} \quad &  \argmin_\mathcal{S} \sum\limits_{\substack{i=-\frac{a-1}{2}\\ i \neq 0}}^{\frac{a-1}{2}}\sum\limits_{\substack{{k=-\frac{l-1}{2}}\\ k \neq 0}}^{\frac{l-1}{2}} \mu_{k,i} \mathcal{J}_{TD}^{x,z}(k,i, \mathcal{S})=\mathcal{T} \\
\end{aligned}
\end{equation}
\noindent with
\begin{equation}
    \mathcal{J}_{TD}^{x,z}(k,i, \mathcal{T}) = \mathscr{L}_{TD}(x+k,z+i|\Delta x, \mathcal{T})
\end{equation}

Finding the best values of $\mu_{k,i}$ is very difficult as the spatial coupling varies from tissue to tissue. A simple and logical workaround can be to use the concept of particles closer to the center having greater relevance. For this, a 2D weighted kernel is adopted that provides spatial coupling, depicted in figure \ref{gaussian_win}. The kernel size is $l \times a$ shaped and generated using a discrete Gaussian function. The mentioned kernel, $g_{l,a}^{\sigma_w}(x,z)$, and therefore, $\mu_{k,i}$ weights are defined as

\begin{equation}
w(x,z) = \frac{1}{2\pi\sigma_w^2} exp\bigg(-\frac{x^2+z^2}{2\sigma_w^2}\bigg)
\end{equation}
\begin{equation}
z \in \bigg[-\frac{a-1}{2}, \frac{a-1}{2}\bigg]; \; x \in \bigg[-\frac{l-1}{2}, \frac{l-1}{2}\bigg]
\end{equation}
where the $\sigma_w$ determines the spatial spread of the Gaussian function.
\begin{equation} \label{gauss}
\mu_{x,z} = g_{l,a}^{\sigma_w}(x,z) = \frac{w(x,z)}{\sum_{z=-\frac{a-1}{2}}^{z=\frac{a-1}{2}}\sum_{x=-\frac{l-1}{2}}^{x=\frac{l-1}{2}} w(x,z)}
\end{equation}
Here, the normalization shown in (\ref{gauss}) is done so that $\sum_{z=-a/2+1}^{a/2-1}\sum_{x=-l/2+1}^{l/2-1} g_{l,a}^{\sigma_w}(x,z) = 1.0$ is maintained. The particle weighted by $\mu_{0,0} = g_{l,a}^{\sigma_w}(0,0)$ shall obtain the highest factor, as it will cover the particle in observation and contribute to the objective function. The shifting of signal groups will be given less weight the further away the corresponding particles are from the center. 
The constrained optimization utilizing (\ref{constr_opt_LTD}) will provide the time-domain-based best shift parameter, $\mathcal{T}_{TD}$.


\begin{figure}[h]
    \centering
    \centerline{\includegraphics[width=0.48\textwidth]{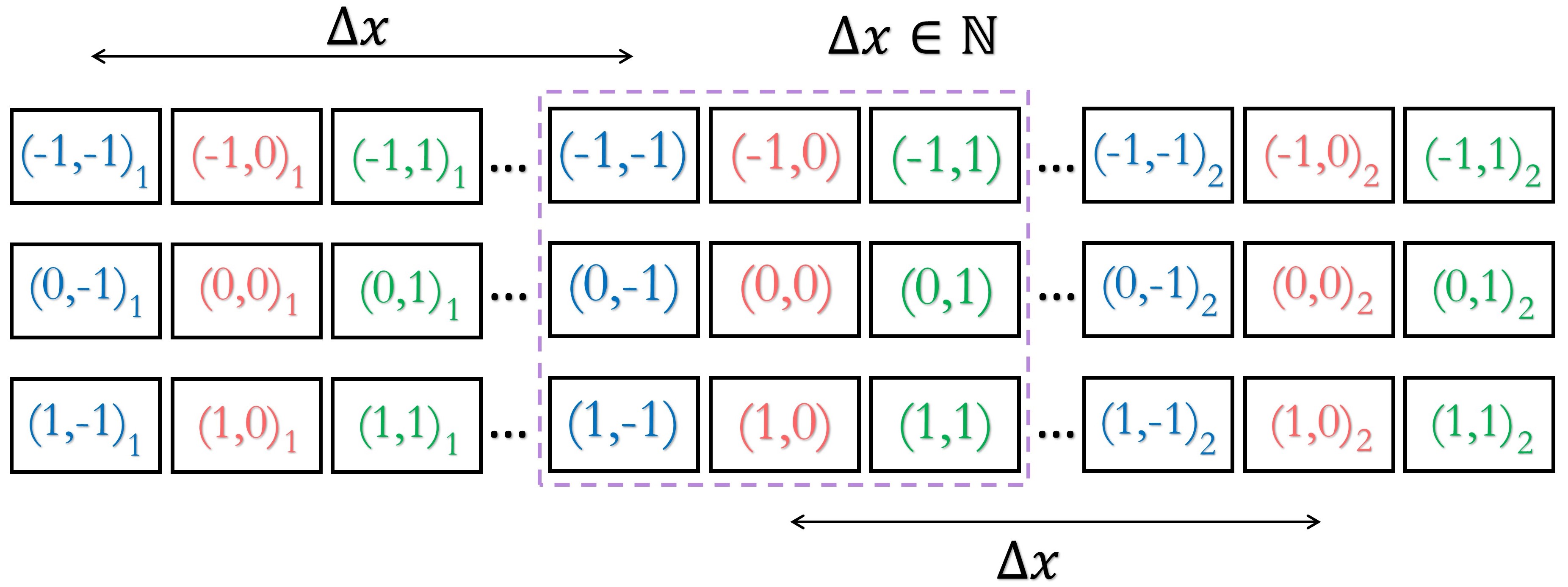}}
    \caption{A $(3\times 3)$ neighborhood (dotted box) and all the respective signal groups $(k,i)_j$ of the particles $(k,i)$ within the kernel.}
    \label{total_win_sigs}
\end{figure}

In figure \ref{total_win_sigs}, an $(l \times a)=(3\times 3)$ local region is considered to demonstrate how $\mathcal{T}_{TD}$ is obtained. Each particle within the neighborhood (depicted as $(k,i)$) can be viewed to produce a $u_0(n)$ signal. Therefore, each of them will have a $u_1(n)$ found from a distance of $\Delta x$ on the $-x$ side and $u_2(n)$ found on the $+x$ side. These are shown as $(k,i)_1$ and $(k,i)_2$ in figure  \ref{total_win_sigs}. If the distance $\Delta x$ covers a homogeneous region, all these signals should have nearly the same shift parameter. Since one particle contributes $2$ shift parameters, a $(l \times a)$ region will contribute a total of $2\times l \times a$ ($18$ for a $3 \times 3$ area, $2$ from the objective function and $16$ from the constraint function). Each signal group's shift, arriving from $(k,i), (k,i)_1, (k,i)_2$ particles, are given the same weight, $\mu_{i,k}$, with their loss term $\mathscr{L}_{TD}(x+k,z+i|\Delta x,\mathcal{T})$. The optimization task minimizes the objective loss function constrained with the cumulative loss of the local region, all of which is parameterized by the shift $\mathcal{T}$. A single shift parameter $\mathcal{T}_{TD}$ can perform the best constrained optimization in the defined $(l \times a)$ area and thus determines the regional shear wave arrival time.

This shift, $\mathcal{T}_{TD}$, acquired from the optimization will be representative of signals sampled at a frequency of $L$ times higher than before due to signal interpolation. Therefore, the SWS in observation can be calculated as

\begin{equation} \label{SWS1}
\mathcal{C}_{TD} = \frac{\Delta x}{\mathcal{T}_{TD}} \cdot \frac{(L.F_s)}{F_{sp}} \cdot 10^{-3}\;\; [ms^{-1}]
\end{equation}

\subsection{Phase-Alignment-based MSE for SWS Estimation} \label{phase_mse}

The displacement signals $u_0(n), u_1(n)$ and $ u_2(n)$ in the frequency domain have similar magnitude spectra. However, the unwrapped versions of their phases are not the same because they are not aligned with each other. The $N$-point Discrete Fourier Transform (DFT) of  (\ref{omega_tau_01_discr}) and (\ref{omega_tau_20_discr}) helps in understanding the reason:
\begin{equation} \label{omega_tau_freq_01}
U_0(e^{j\omega}) =  \alpha_{10}\;e^{j\omega\tau} U_1(e^{j\omega});\; \alpha_{10}<1.0
\end{equation}
\begin{equation} \label{omega_tau_freq_20}
U_2(e^{j\omega}) =  \alpha_{02}\;e^{j\omega\tau} U_0(e^{j\omega});\; \alpha_{02}<1.0
\end{equation}
The digital angular frequency $\omega$ has a range between $[-\pi,\pi]$. The $e^{j\omega\tau}$ term misaligned the phase spectra of the displacement signals which should be theoretically equal for both (\ref{omega_tau_freq_01}) and (\ref{omega_tau_freq_20}). The shift term $\tau$ can be determined from the phase components as
\begin{equation} \label{phase_angle_m02}
\begin{split}
\angle U_m(e^{j\omega}) = & \angle \left[ \alpha\;e^{j\omega\tau} U_{1-\frac{m}{2}}(e^{j\omega})\right] \\
 = & \angle \alpha + \angle e^{j\omega\tau}+  \angle U_{1-\frac{m}{2}}(e^{j\omega}); \; m \in [0,2]
\end{split}
\end{equation}
Rearranging (\ref{phase_angle_m02}) and with $\angle \alpha=0$:
\begin{equation}
\begin{split} \label{U_m_angle}
\angle U_m(e^{j\omega})-\angle U_{1-\frac{m}{2}}(e^{j\omega}) \; & = \angle e^{j\omega\tau}= \omega\tau\\
\end{split}
\end{equation}
Then from (\ref{U_m_angle}), $\tau$ can be obtained as
\begin{equation} \label{shift_theory}
 \tau = \frac{\angle U_m(e^{j\omega})-\angle U_{1-\frac{m}{2}}(e^{j\omega})}{\omega}
\end{equation}
Theoretically, any digital frequency should provide the exact shift parameter from (\ref{shift_theory}). This calculation of the shift parameter will work in case of a distortionless and noiseless system. However, in practical scenarios, no two displacement signals (from particles separated by even small distances) possess scalable magnitudes or phases across all frequencies; this is because the system has non-linearity and distortion. The direct utilization of the theoretical relation in (\ref{shift_theory}) cannot be done which some literature overlooks (i.e., FDSM \cite{rosen2018fourier}). 


There are two considerations to deal with in the frequency domain for estimating the shift parameter in practical scenarios. Firstly, real mediums are not distortionless. This means that real signals possess phase components that are not linearly delayed from one another with respect to frequency. Also, their magnitudes cannot be equalized by scaling with a scalar quantity. This is due to the signal oscillations across the temporal axis not being identical. As a result, (\ref{omega_tau_freq_01}) and (\ref{omega_tau_freq_20}) are to be rewritten as

\begin{equation} \label{omega_tau_freq_01_rewritten}
U_0(e^{j\omega}) \cong  \alpha_{10}\;e^{j\omega\tau} e^{j\omega\mathcal{Q}_{10}}U_1(e^{j\omega});\; \alpha_{10}<1.0
\end{equation}
\begin{equation} \label{omega_tau_freq_20_rewritten}
U_2(e^{j\omega}) \cong  \alpha_{02}\;e^{j\omega\tau}e^{j\omega\mathcal{Q}_{02}} U_0(e^{j\omega});\; \alpha_{02}<1.0
\end{equation}
Here, $e^{j\omega\mathcal{Q}_{10}}$ and $e^{j\omega\mathcal{Q}_{02}}$ contribute non-linear phases which cannot be pre-determined; they can be effectively considered as noise. Secondly, phase values become erratic in high-frequency regions of the signals as the significant frequency components lie in a certain range. Small oscillations caused by noise or residual values at the tail end of $u_k(n)$ are responsible for this particular issue. Both of these challenges are shown in figure \ref{phase_0}\textcolor{blue}{(a-d)} with time and frequency domain displacement signals.


Realistic signals from the displacement of equidistant and homogenous particles are used in figure  \ref{phase_0}. Plot  \ref{phase_0}\textcolor{blue}{(a)} shows three displacement signals with unequal peaks in the time domain. After the signals decay out, some small residual oscillations remain at the tail-end, e.g., after $8$~$ms$. The associated unwrapped phases show non-linearity in the low-frequency range and prominent noisy effects in the high-frequency band ($>1000 Hz$, figure  \ref{phase_0}\textcolor{blue}{(b)}), but have very low magnitude (figure  \ref{phase_0}\textcolor{blue}{(d)}). This results in the phase differences between $U_1(e^{j\omega})$ and $U_0(e^{j\omega})$ $($i.e., $\angle U_0(e^{j\omega})-\angle U_1(e^{j\omega}))$ as well as $U_0(e^{j\omega})$ and $U_2(e^{j\omega})$ $($i.e., $\angle U_2(e^{j\omega})-\angle U_0(e^{j\omega}))$ will demonstrate non-linearity and noisy components (figure  \ref{phase_0}\textcolor{blue}{(c)}) across the frequency axis. However, the temporal delays of the mentioned signals with respect to one another should be equal. Therefore, according to (\ref{shift_theory}), both $\angle U_0(e^{j\omega})-\angle U_1(e^{j\omega})$ and $\angle U_2(e^{j\omega})-\angle U_0(e^{j\omega})$ should be equal as well for any $\omega$.

In contrast, three signals are visualized in figure  \ref{phase_0}\textcolor{blue}{(e)} which have been obtained by cleaning the ones in figure \ref{phase_0}\textcolor{blue}{(a)} to have equal peaks (normalized) as well as zeroed-out tails. As can be seen from figure \ref{phase_0}\textcolor{blue}{(f)}, the corresponding phases are better distinguishable than before. At low frequencies, the phase curves are close to a straight line and the non-linear phases ($\angle e^{j\omega\mathcal{Q}_{10}}$ and $\angle e^{j\omega\mathcal{Q}_{02}}$) effectively do not exist. If the corresponding phase differences are observed in figure \ref{phase_0}\textcolor{blue}{(g)}, it is found that $\angle U_0(e^{j\omega})-\angle U_1(e^{j\omega})$ and $\angle U_2(e^{j\omega})-\angle U_0(e^{j\omega})$ are nearly equal upto a certain frequency ($1500Hz$), after which they show a slight non-linear deviation from each other. This is because a higher degree of noise components exist in the high-frequency range rather than useful signals. Therefore, the phase lines are much more reliable for cleaned signals in lower frequency ranges compared to the higher range. Speculating the normalized magnitude spectra in figure \ref{phase_0}\textcolor{blue}{(h)}, the harmonic components with less than $0.178\;(i.e., -7.5$dB) times less magnitude of the most prominent component reside outside $1500Hz$, consisting to the non-linearity in the phase lines. A significant frequency component ($F_{sig}<F_s/2$) can be chosen to isolate the low and high-frequency components. Nevertheless, the approximate linearity of phase lines and equality of the required phase differences necessitate the cleaning of displacement signals. 


The claim that the phase differences $(\angle U_0(e^{j\omega}) - \angle U_1(e^{j\omega}))$ and $(\angle U_2(e^{j\omega}) - \angle U_0(e^{j\omega}))$ are straight lines can be made assuming the tissue medium is elastic, isotropic, and possesses a linear relationship between stress and strain, based on section \ref{problem_form_section}. But, as explained in the previous paragraph, there exists some degree of non-linearity even within the significant frequency range (figure \ref{phase_0}\textcolor{blue}{(g)}) which cannot be fully eliminated from signal cleaning. Thus, straight lines are fitted to $(\angle U_0(e^{j\omega}) - \angle U_1(e^{j\omega}))$ and $(\angle U_2(e^{j\omega}) - \angle U_0(e^{j\omega}))$ using linear-regression with respect to the frequency axis to impose linearity. The intercept parameters from the line-fitting are ignored for convenience as the slope parameters fundamentally contribute to the signal shift. Any points above the selected significant frequency range, $F_{sig}$, are discarded. The two regressed straight lines are then divided by the slope of $\angle e^{j\omega}$ (which is $\omega$, according to (\ref{shift_theory})). To obtain the shift parameter, this entire process is mathematically described below:
\begin{equation} \label{reg_phase_01}
m_{01}, \lambda_{01} \leftarrow Reg^1 \left[\angle U_0(e^{j\omega}) - \angle U_1(e^{j\omega})\right] \bigg|_{\omega_1 \le \omega \leq  \omega_{sig}}
\end{equation}
\begin{equation} \label{reg_phase_20}
 m_{20}, \lambda_{20} \leftarrow Reg^1 \left[\angle U_2(e^{j\omega}) - \angle U_0(e^{j\omega})\right] \bigg|_{\omega_1 \le \omega \leq  \omega_{sig}}
\end{equation}
\begin{equation}
\angle U_{01}(e^{j\omega}) = m_{01}\omega;\; \angle U_{20}(e^{j\omega}) = m_{20}\omega
\end{equation}

\begin{equation} \label{tau_phase01}
\tau_{01}  = \frac{1}{P} \sum_{\omega= \omega_1}^{\omega_{sig}} \frac{\angle U_{01}(e^{j\omega})}{\omega}
\end{equation}
\begin{equation} \label{tau_phase20}
\tau_{20}  = \frac{1}{P} \sum_{\omega= \omega_1}^{\omega_{sig}} \frac{\angle U_{20}(e^{j\omega})}{\omega} 
\end{equation}
Here, $Reg^1[\cdot]$ is used to indicate the 1st-degree polynomial regression (line-fit). The quantities $m$ and $\lambda$ are corresponding slope and intercept parameters, respectively, from regression. The DC-frequency, $\omega=0$ is ignored. This is because DC components provide no phase difference $(\angle U_{01}(0)=\angle U_{20}(0)=0)$ and including them would result in zero over zero division. As such, the first frequency index is considered to be $\omega_1 = \frac{2\pi F_s}{N}$, avoiding $\omega=0$, as the starting of the regression range and $\omega_{sig} = \frac{2\pi F_{sig}}{F_s}$ as the ending. Thus, both the line fitted phase and $\omega$ shall have $P=\frac{NF_{sig}}{2F_s}-1$ points. The shift parameters are finally calculated by averaging across $P$ frequency points, as shown in (\ref{tau_phase01}) and (\ref{tau_phase20}).

In practice, $\tau_{01}$ and $\tau_{20}$ may not be equal. And, being calculated from a division, the two shift parameters will have fractional values. Fractional delays in DT signals (with finite samples) are irrational unless the signals are interpolated. None of the signals were interpolated prior to entering the frequency domain. As such, the interpolation factor $(L)$ can be introduced at this stage as follows:
\begin{equation} \label{T01_phase}
\mathcal{T}_{01}  = interp \big[\tau_{01}.L \big]
\end{equation}
\begin{equation} \label{T20_phase}
\mathcal{T}_{20}  = interp \big[\tau_{20}.L\big]
\end{equation}
The $interp[\cdot]$ operation indicates conversion to an integer value. It is to be noted that when $L=1$, the resulting shift parameters are representative of the original signals $u_0(n), u_1(n),$ and $u_2(n)$ with no interpolation.

\begin{table*}[h]
\centering
\normalsize

\begin{tabular}{l}

\hline \\

Using (\ref{constr_opt_LTD}) and (\ref{constr_opt_LPD}), the combined objective and constraint functions can be obtained as\\ 


\end{tabular}
\end{table*}

\begin{figure*}[h]
    \centering
    \centerline{\includegraphics[width=0.99\textwidth]{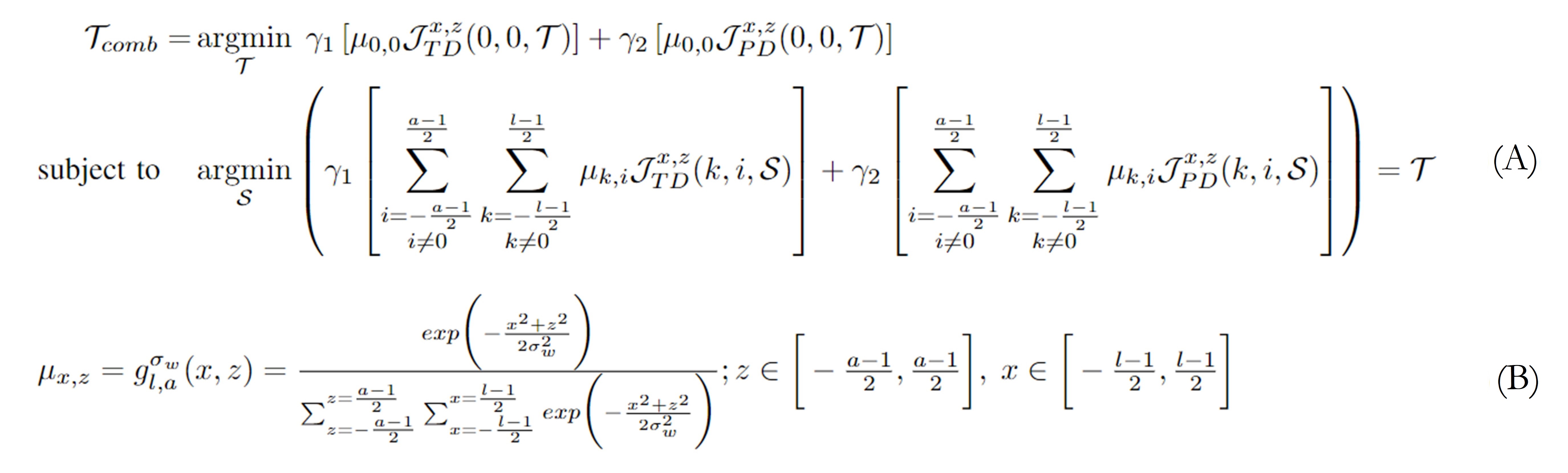}}
\end{figure*}

\begin{table*}[h]
\centering
\normalsize

\begin{tabular}{l}
\textcolor{white}{Using (ab) and (cd), the combined objective and constraint functions can be obtained as} \\
\hline
\end{tabular}
\end{table*}

Now, the stage has been reached to formulate the constrained optimization problem for phase alignment. Unlike the time-domain approach where one of the signals is iteratively shifted to procure the best possible NCC with another signal, this method utilizes $\angle e^{j\omega}$ and iteratively changes its slope to get the best alignment of two-phase differences. Furthermore, components with higher amplitude in the magnitude spectrum provide phases that transition more linearly than lower amplitude components. For this reason, the phase alignment is best weighted using the normalized magnitude spectrum. A loss function is initially designed to take all these considerations into account for a particle in observation ($PD$ subscript indicates phase domain): 
\begin{equation}
|U(e^{j\omega})| = \frac{1}{3} \sum_{k=0}^{2} |U_{k}(e^{j\omega})| \bigg|_{\omega_1 \le \omega \leq  \omega_{sig}}
\end{equation}
\begin{equation}
|U| = \frac{|U(e^{j\omega})|}{max\big[|U(e^{j\omega})|\big]+ \varepsilon}
\end{equation}
\begin{equation} \label{01_mse}
\mathscr{L}_{01,PD}(x,z|\Delta x,\mathcal{T}) = \frac{1}{\mathrm{P}} \sum_{\omega= \omega_1}^{\omega_{sig}} \left[\left(\angle U_{01}(e^{j\omega}) -\frac{\mathcal{T}}{L}\omega \right)|U|\right]^2 
\end{equation}
\begin{equation} \label{20_mse}
\mathscr{L}_{20,PD}(x,z|\Delta x,\mathcal{T}) = \frac{1}{\mathrm{P}} \sum_{\omega= \omega_1}^{\omega_{sig}} \left[\left(\angle U_{20}(e^{j\omega})- \frac{\mathcal{T}}{L}\omega \right)|U|\right]^2
\end{equation}
\begin{equation}
\begin{split}
\mathscr{L}_{PD}(x,z|\Delta x,\mathcal{T}) = \; & \mathscr{L}_{01,PD}(x,z|\Delta x,\mathcal{T})\\
& +\mathscr{L}_{20,PD}(x,z|\Delta x,\mathcal{T})
\end{split}
\end{equation}
The shift parameter $\mathcal{T}$ (an integer) is used to control the slope $\mathcal{T}/L$. The MSE calculation in (\ref{01_mse}) and (\ref{20_mse}) form this phase-alignment approach into a minimization problem. The contribution of the center and local particles to the arrival time estimation is similar to the previous section now that the loss function has been devised. The constrained optimization for phase alignment is described below:


\begin{equation} \label{constr_opt_LPD}
\begin{aligned}
\mathcal{T}_{PD} = &\argmin_{\mathcal{T}} \quad  \mu_{0,0} \mathcal{J}_{PD}^{x,z}(0,0, \mathcal{T})\\
\textrm{s.t.} \quad &  \argmin_\mathcal{S} \sum\limits_{\substack{i=-\frac{a-1}{2}\\ i \neq 0}}^{\frac{a-1}{2}}\sum\limits_{\substack{{k=-\frac{l-1}{2}}\\ k \neq 0}}^{\frac{l-1}{2}} \mu_{k,i} \mathcal{J}_{PD}^{x,z}(k,i, \mathcal{S})=\mathcal{T} \\
\end{aligned}
\end{equation}
\noindent with
\begin{equation}
    \mathcal{J}_{PD}^{x,z}(k,i, \mathcal{T}) = \mathscr{L}_{PD}(x+k,z+i|\Delta x, \mathcal{T})
\end{equation}
Here, $\mu_{k,i}$ values are set in the same way from a spatially decaying $g_{l,a}^{\sigma_w}$ kernel. The optimum shift parameter acquired from this task can be depicted as $\mathcal{T}_{PD}$. Finally, the SWS of the centre particle is calculated similarly to (\ref{SWS1}) as 

\begin{equation} \label{SWS2}
\mathcal{C}_{PD} = \frac{\Delta x}{\mathcal{T}_{PD}} \cdot \frac{(L.F_s)}{F_{sp}} \cdot 10^{-3} \;\; [ms^{-1}]
\end{equation}

\subsection{Loss-Combined Constrained Optimization}
The constrained optimization can be designed by combining the losses to obtain both of the benefits from the two domain-based approaches. If any one of the loss functions produces outlier estimations, the other loss can mitigate the error. This combination also provides an ensemble-type effect and isolates the best shift parameter better. Since the losses are different (NCC and MSE), their ranges will vary. Therefore, the objective and constraint functions from the two domains are weighted ($\gamma_1, \gamma_2$) before being combined. The entire loss-combined constrained optimization is given by (\textcolor{blue}{A}) and (\textcolor{blue}{B}).

\subsection{Accelerating Convergence}


For each particle under the $g_{l,a}^{\sigma_w}$ kernel, there are two different corresponding shift parameters: (i) Time NCC: $\mathcal{T}_{10}$ and $\mathcal{T}_{02}$ from (\ref{T01_time}) and (\ref{T20_time}); (ii) Phase MSE: $\mathcal{T}_{10}$ and $\mathcal{T}_{02}$ from (\ref{T01_phase}) and (\ref{T20_phase}). As a result, for each constrained optimization approach, there will be a total of $(2 \times l \times a )$ delays which can be calculated prior to the optimized prediction. The lowest and highest parameters from this particular set can be treated as the iteration range inside which the corresponding losses can be minimized. Being an iterative process for estimating the entire 2D SWS mapping, $\Upsilon(x,z)$, this particular range can accelerate the convergence noticeably.

\section{Materials and Experimental Setup}
\subsection{Simulation Phantom Datasets}

\begin{figure*}[h]
    \centering
    \centerline{\includegraphics[width=0.98\textwidth]{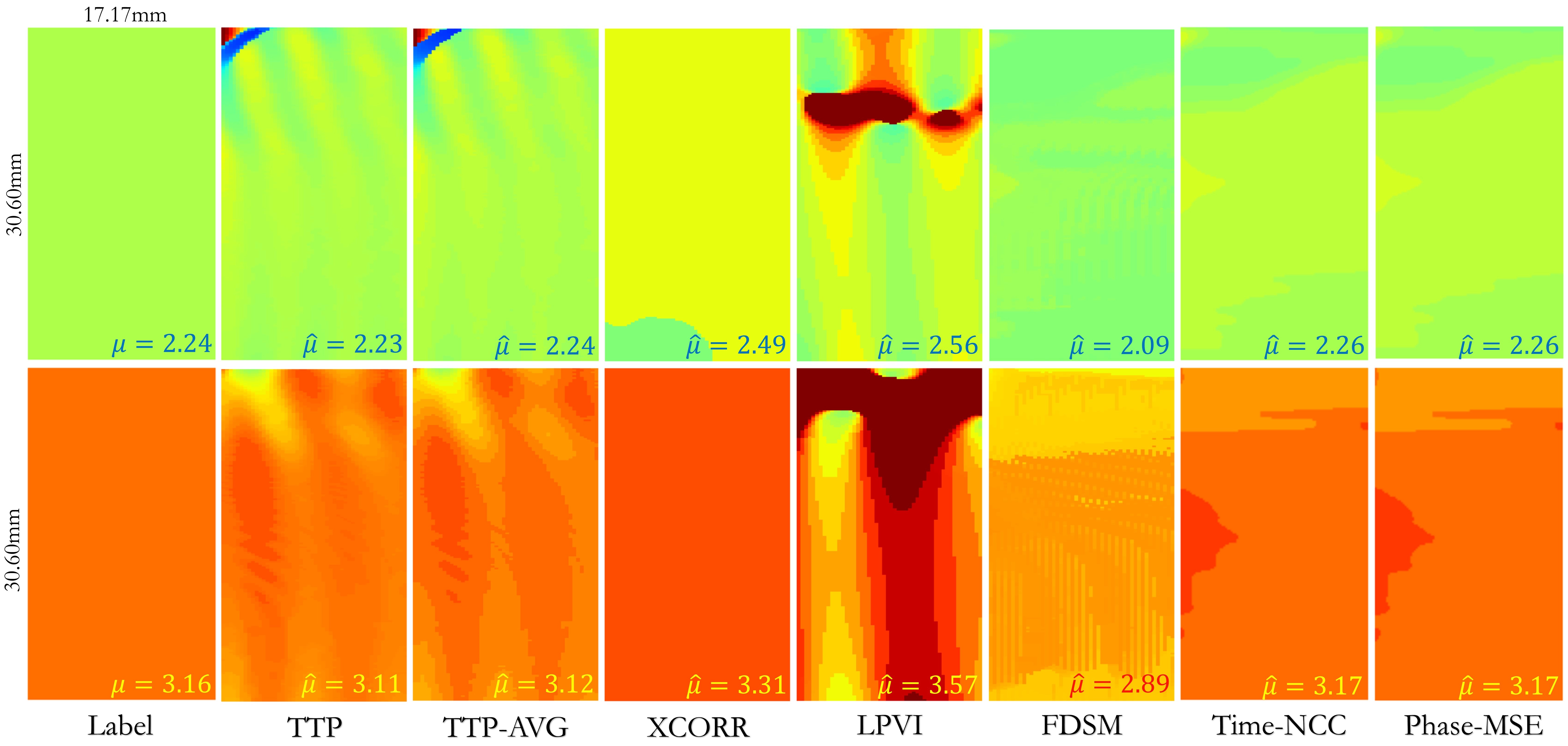}}
    \caption{Reconstruction outcomes from simulation data with the proposed techniques and the state-of-the-art methods. Here, $\mu$ indicates mean value of SWS.}
    \label{homogenous_phantom}
\end{figure*}

Simulation phantom data was obtained from the open repository of the Duke University research repository, RSNA Quantitative Imaging Data Warehouse (US-SWS-Digital-Phantoms) \cite{simu_data}. The data contains displacement fields produced from homogenous elastic mediums. Each phantom data was generated by the Finite-Element Modeling of ARF \cite{palmeri2016guidelines} using the Field II software package. The displacement fields, $u(x,z,t)$, have varying time-frames for different data instances; but they possess at least $70$ frames depicting SW propagation. The homogeneous mediums of $15kPa$ and $30kPa$ are used from these phantoms to demonstrate the accuracy of the proposed optimization approaches.   

\subsection{Experimental Phantom Datasets}
\subsubsection{Data-1}
The first experimental dataset is provided by the Ultrasound Research Laboratory of the Mayo Clinic's Department of Radiology, USA (CIRS Phantom Model 049). It contains $4$ different circular inclusion sizes with $2$ varieties of stiffness with respect to the background, making a total of $8$ distinct data samples. A single ARF push is provided to produce shear waves which are tracked and processed using a 2D auto-correlating Loupas Algorithm \cite{loupas1995experimental, loupas1995axial}. The ARF is provided at the laterally left side of the inclusions. The data contains the ground labels of the inclusion and background. The structural specifications of the samples are provided in table \ref{table:model049_specs}. 
\begin{table}[h]
\centering
\caption{Structural information of Data-1}
\footnotesize
\label{table:model049_specs}
\begin{tabular}{crrr}
\hline
{Data$-$Sample}  &\begin{tabular}[c]{@{}c@{}}{Diameter}\\$[mm]$\end{tabular}   & \begin{tabular}[c]{@{}c@{}}{Inclusion}\\ {Stiffness,} $[kPa]$\end{tabular}  & \begin{tabular}[c]{@{}c@{}}{Background}\\ {Stiffness,} $[kPa]$\end{tabular} \\ \hline
$1-1$   &  $10.40$ & $45.00$ & $25.00$\\ 
$1-2$  &  $6.49$  & $45.00$ & $25.00$\\ 
$1-3$  &  $4.05$  & $45.00$ & $25.00$\\ 
$1-4$  &  $2.53$  & $45.00$  & $25.00$ \\ 
$1-5$  &  $10.40$  & $80.00$ & $25.00$\\ 
$1-6$   &  $6.49$   & $80.00$ & $25.00$\\ 
$1-7$  &  $4.05$   & $80.00$ & $25.00$\\ 
$1-8$  &  $2.53$   & $80.00$  & $25.00$\\ \hline
\end{tabular}
\end{table}


\subsubsection{Data-2}
This data has been collected from the Duke University research repository (RSNA-QIBA-US-SWS) \cite{data2}. It contains IQ data of eight elastic phantoms. The SWS of the phantoms were estimated using a Vantage research scanner sequence (Verasonics, Inc, Kirkland, WA). The ARF is given in the middle of the phantom \cite{deng2016ultrasonic}. The nominal SWS is reported to be within $1-2$~$ms^{-1}$ for all the phantoms. 

\subsubsection{Data-3}
This private dataset has been acquired with an ultrasound system equipped with an $L64$ linear array transducer. The ARF pulse and Tracking pulse were both set to be $5MHz$. The data comprises four experimental CIRS model 049 phantoms. It contains $4$-region-based SWE for each phantom where every region is faced with isolated ARF pulses. The regions physically overlap each other. With an ARF push to the right side of the ROI, the tracked data has been processed and directionally filtered to obtain the final displacement map. The stiffness information of the samples used in the result generation is shown in table \ref{table:private_specs}.

\begin{table}[h]
\centering
\caption{Structural information of Data-3}
\footnotesize
\label{table:private_specs}
\begin{tabular}{ccc}
\hline
{Data$-$Sample}  &\begin{tabular}[c]{@{}c@{}}{Inclusion}\\{Stiffness,} $[kPa]$\end{tabular}   & \begin{tabular}[c]{@{}c@{}}{Background}\\ {Stiffness,} $[kPa]$\end{tabular}  \\ \hline

$3-1$  &  $9.00$   & $20.00$ \\ 
$3-2$  &  $76.00$   & $21.00$  \\ \hline
\end{tabular}
\end{table}

\begin{table}[h]
\centering
\caption{Specifications of the Datasets}
\footnotesize
\label{table:all_data_specs}
\begin{tabular}{cccc}
\hline
{Data} &\begin{tabular}[c]{@{}c@{}}{$F_s$}\\$[kHz]$\end{tabular}  &\begin{tabular}[c]{@{}c@{}}{$R_s$}\\$[mm/pix]$\end{tabular}   & \begin{tabular}[c]{@{}c@{}}{Dimension}\\$(X,Z,T)$\end{tabular}\\ \hline
$Simu$ & $10.0$  &  $l,a=0.17$ & $(150, 599, 70+)$\\ 
$1$ & $11.765$  &  $l,a=0.154$ & $(300, 300, 178)$\\ 
$2$ & $5.0$  &  $l=0.2, a=0.246$  & $(475, 436, 200)$\\ 
$3$ & $8.0$  &  $l=0.163, a=0.128$  & $(43, 201, 304)$ \\ \hline

\end{tabular}
\end{table}

\begin{table*}[h]
\centering
\caption{Quantitative comparison of different estimation techniques for Data-1 and Data-3}
\footnotesize
\label{table:quantitative}
\begin{tabular}{llccccccc}
\hline
Samples & Metrics              & TTP             & TTP-Avg        & XCorr           & LPVI            & FDSM            & NCC             & MSE             \\ \hline
1-1          & $\hat{\mu}_I \pm \hat{\sigma}_I$ & 4.06$\pm$0.51 & 4.2$\pm$0.49 & 3.39$\pm$0.4  & 3.32$\pm$0.29 & 3.53$\pm$0.21 & 3.94$\pm$0.28 & 3.81$\pm$0.27 \\
           & $\hat{\mu}_B \pm \hat{\sigma}_B$ & 3.12$\pm$0.42 & 3.21$\pm$0.4 & 2.63$\pm$0.33 & 2.53$\pm$0.32 & 2.81$\pm$0.81 & 2.96$\pm$0.24 & 2.89$\pm$0.22 \\
                            & $\mathrm{PSNR} (dB)$          & 10.75           & 13.03          & 9.73            & 14.42           & 22.75           & 18.91           & 17.69           \\
                            & $\mathrm{CNR} (dB)$           & 6.9             & 8              & 7.39            & 7.88            & 11.81           & 12.39           & 12.32           \\ \hline
1-2          & $\hat{\mu}_I \pm \hat{\sigma}_I$ & 3.83$\pm$0.57  & 3.95$\pm$0.54 & 3.16$\pm$0.4 & 3.07$\pm$0.18 & 3.34$\pm$0.25 & 3.64$\pm$0.27 & 3.52$\pm$0.26 \\
            & $\hat{\mu}_B \pm \hat{\sigma}_B$ &  3.05$\pm$0.37  & 3.13$\pm$0.33 & 2.57$\pm$0.32  & 2.55$\pm$0.27 & 2.77$\pm$0.83 & 2.91$\pm$0.16 & 2.84$\pm$0.16 \\
                            & $\mathrm{PSNR} (dB)$          &     12.41            &     13.01           &     9.84            &        16.54         &     23.13            &    18.39             &  17.3 
                            \\
                            & $\mathrm{CNR} (dB)$           &      6.36           &      7.77          &      5.36           &      5.77           &       10.46          &     12.83            &   12.36         \\ \hline
1-3          & $\hat{\mu}_I \pm \hat{\sigma}_I$ & 3.45$\pm$0.38 & 3.55$\pm$0.38 & 2.87$\pm$0.34 & 2.86$\pm$0.17 & 3.0$\pm$0.2 & 3.27$\pm$0.22 & 3.18$\pm$0.2 \\
            & $\hat{\mu}_B \pm \hat{\sigma}_B$ &  3.04$\pm$0.34 & 3.11$\pm$0.3 & 2.53$\pm$0.29 & 2.57$\pm$0.15 & 2.76$\pm$0.85 & 2.88$\pm$0.13 &  2.81$\pm$0.13 \\
                            & $\mathrm{PSNR} (dB)$          &        14.74   &   18.27   &    9.91    &   12.33         &     24.85         &     22.57     &    23.39             \\
                            & $\mathrm{CNR} (dB)$           &      1.91    &       3.48     &       1.16    &   5.5              &   6.55         &      9.38        &      9.02       \\ \hline
1-4          & $\hat{\mu}_I \pm \hat{\sigma}_I$ &  3.08$\pm$0.27  &    3.17$\pm$0.22 &  2.56$\pm$0.18    &  2.63$\pm$0.14    & 2.79$\pm$0.06  &    2.96$\pm$0.1    &     2.88$\pm$0.09   \\
            & $\hat{\mu}_B \pm \hat{\sigma}_B$ &  3.03$\pm$0.32    &    3.11$\pm$0.28     &   2.52$\pm$0.23    & 2.58$\pm$0.24  &  2.78$\pm$0.85  &   2.88$\pm$0.11 &  2.81$\pm$0.1    \\
                            & $\mathrm{PSNR} (dB)$          &      13.1        &      16.15      &   10     &       13.41     &     15.71      &     28.58     &      28.76      \\
                            & $\mathrm{CNR} (dB)$           &      -15.02  &      -12.58  &   -14.13   &   -14.56          &    -18.63     &     -2.55      &     -2.57        \\ \hline
1-5          & $\hat{\mu}_I \pm \hat{\sigma}_I$ &  5.37$\pm$0.52  &   5.51$\pm$0.45    &    4.72$\pm$0.5  &       
            4.41$\pm$0.53       &     4.17$\pm$0.09    &   5.52$\pm$0.19    &   5.28$\pm$0.24              \\
           & $\hat{\mu}_B \pm \hat{\sigma}_B$ &  3.45$\pm$0.81   &   3.56$\pm$0.82    &    2.93$\pm$0.67   &  2.62$\pm$0.54   &  3$\pm$0.77  &    3.31$\pm$0.75   &    3.22$\pm$0.71     \\
                            & $\mathrm{PSNR} (dB)$          &    17.26    &     17.12    &    16.99    &  18.18         &     18.33         &    18.35     &    18.7     \\
                            & $\mathrm{CNR} (dB)$           &   7.49   &     7.61  &    8.5    &    10.39        &       7.81          &     9.06      &     9.06         \\ \hline
1-6          & $\hat{\mu}_I \pm \hat{\sigma}_I$ & 5.01$\pm$0.55 &  5.20$\pm$0.50 &  4.22$\pm$0.47   &  3.97$\pm$0.32              &  3.97$\pm$0.16   &   5.17$\pm$0.37   &   5.16$\pm$0.36         \\
           & $\hat{\mu}_B \pm \hat{\sigma}_B$ &   3.45$\pm$0.81    &   3.56$\pm$0.82   &  2.93$\pm$0.67    &  2.62$\pm$0.54   &  3.00$\pm$0.77  &   3.26$\pm$0.41   &  3.25$\pm$0.40  \\
                            & $\mathrm{PSNR} (dB)$          &    20.17    &   20.57    &   17.79    &  19.49     & 18.95          &     22.23         &    22.61             \\
                            & $\mathrm{CNR} (dB)$           &   10.36     &    11.06    &    11.00    &  11.44        &     11.44     &     13.33       &     13.26    \\ \hline
1-7          & $\hat{\mu}_I \pm \hat{\sigma}_I$ & 4.59$\pm$0.53 &  4.75$\pm$0.51   &  3.80$\pm$0.44    &          
            3.56$\pm$0.2      &    3.84$\pm$0.17   &   4.23$\pm$0.25   &   4.26$\pm$0.23    \\
            & $\hat{\mu}_B \pm \hat{\sigma}_B$ &  3.10$\pm$0.40 &   3.18$\pm$0.37   &   2.61$\pm$0.30  &  2.60$\pm$0.24  &   2.82$\pm$0.80   &   2.96$\pm$0.26   &    2.88$\pm$0.25  \\
                            & $\mathrm{PSNR} (dB)$          &    22.06    &    23.10   &    17.66    &  22.04               &     19.76     &     23.69     &    22.83             \\
                            & $\mathrm{CNR} (dB)$           &    11.45     &    12.42      &   11.92  &  12.20               &    14.48        &     15.10     &    15.02      \\ \hline
1-8          & $\hat{\mu}_I \pm \hat{\sigma}_I$ & 4.07$\pm$0.53 & 4.19$\pm$0.49  &  3.35$\pm$0.38  &  3.17$\pm$0.22                         &   3.40$\pm$0.36    &     3.95$\pm$0.37      &   3.83$\pm$0.34      \\
           & $\hat{\mu}_B \pm \hat{\sigma}_B$ & 3.06$\pm$0.32 &  3.14$\pm$0.30 &  2.54$\pm$0.23  &  2.58$\pm$0.18               &  2.80$\pm$0.82      &    2.91$\pm$0.17   &   2.85$\pm$0.16              \\
                            & $\mathrm{PSNR} (dB)$          &   23.84    &   21.91    &   20.36     &  17.78           &    16.21   &    23.20     &     22.12           \\
                            & $\mathrm{CNR} (dB)$           &   9.99    &   10.97   &    11.01    &   10.44              &     13.82           &     16.06            &      16.00           \\ \hline
3-1 & $\hat{\mu}_I \pm \hat{\sigma}_I$ & 1.66$\pm$0.10 & 1.67$\pm$0.10 & 1.82$\pm$0.15 & - & 1.44$\pm$0.05 & 1.71$\pm$0.11& 1.77$\pm$0.11 \\
& $\hat{\mu}_B \pm \hat{\sigma}_B$ & 1.99$\pm$0.27 & 2.00$\pm$0.31 &2.29$\pm$0.43 & - & 1.68$\pm$0.17 & 2.07$\pm$0.21 & 2.13$\pm$0.21 \\ 
&$\mathrm{PSNR} (dB)$ & 18.61 & 19.45 & 23.93 & - & 25.28 & 17.95 & 18.82 \\
& $\mathrm{CNR} (dB)$ & 1.66 & 0.79 & 0.76 & - & 2.91 & 4.69 & 4.55 \\ \hline

3-2 & $\hat{\mu}_I \pm \hat{\sigma}_I$ & 3.21$\pm$0.62 & 3.23$\pm$0.60 & 3.94$\pm$0.60 & - & 2.34$\pm$0.22 & 3.29$\pm$0.41 & 3.24$\pm$0.42 \\
& $\hat{\mu}_B \pm \hat{\sigma}_B$ & 2.30$\pm$0.41 & 2.34$\pm$0.46 & 2.83$\pm$0.65 & - &  1.96$\pm$0.11 & 2.38$\pm$0.35 & 2.35$\pm$0.34 \\
& $\mathrm{PSNR} (dB)$  & 13.9 & 14.17 & 16.11 & - & 14.96 & 18.20 & 18.17 \\
& $\mathrm{CNR} (dB)$ & 6.77 & 5.76 & 4.63 & - & 10.80 & 8.22 & 8.21 \\ \hline
   
\end{tabular}
\end{table*}

The specifications (temporal frequency, $F_s$, lateral resolution, $R_s$, and data dimension) of each data are provided in Table \ref{table:all_data_specs}. The simulation phantoms are depicted as `$Simu$'.

\subsection{Performance Metrics}

The performance evaluation of the proposed approaches and the state-of-the-art methods is observed based on the following metrics: (i) Mean and Standard-Deviation of the estimated 2D SWS of inclusion ($\hat{\mu}_I, \hat{\sigma}_I$) and background ($\hat{\mu}_B, \hat{\sigma}_B$) regions, (ii) Contrast-to-Noise Ratio ($\mathrm{CNR}$), and (iii) Peak-Signal-to-Noise-Ratio ($\mathrm{PSNR}$). The latter two are described below:

\textbf{CNR:} $\mathrm{CNR}$ is calculated using quantities (i.e., mean and standard-deviation) from the estimated inclusion ($I$) and background ($B$), separately:
    \begin{equation} 
\mathrm{CNR} = 20\;log_{10} \bigg(\frac{\big| \hat{\mu}_I - \hat{\mu}_B \big|}{ \sqrt{\hat{\sigma}_I^2+\hat{\sigma}_B^2}
 }\bigg) \;dB
\end{equation}

\textbf{PSNR:} The mean-squared-error ($\mathrm{mse}$) of the normalized reconstruction ($\Upsilon$) with respect to the normalized label ($\Upsilon_{l}$) is used to calculate the corresponding $\mathrm{PSNR}$:
\begin{equation} 
\mathrm{mse} = \left| \left| \frac{\Upsilon_{l}}{max(\Upsilon_l)}-\frac{{\Upsilon}}{max({\Upsilon})}  \right| \right|_2
\end{equation}
\begin{equation} 
\mathrm{PSNR} = 10\;log_{10}\bigg(\frac{1.0}{\mathrm{mse}}\bigg)\;dB
\end{equation}    

\subsection{Experimental Setup} \label{exp_setup}

The parameters related to the optimization (figure  \ref{GA}) influence the SWS map reconstruction of each phantom significantly. The choice of parameters in this study for obtaining optimum results is provided below:

\begin{itemize}
    \item $M=2$
    \item $T_{sh} =250, \; q=0.9, \; \varrho=1, \; r=3$
    \item $\Delta x=0.5$~$mm$ ($Simu$, Data-3), $1.0$~$mm$ (Data-1, 2)$, \; (a,l)=5,\; \sigma_w=1, \; L=10, \; F_{sig}=500Hz\;$(Data-2) or $1000Hz\;$($Simu$, Data-1, 3)
    \item Post-Processing: Median Filter with $(5\times 5)$ window size
\end{itemize}




Since the metrics require ground labels, quantitative results corresponding to only Data-1 and Data-3 are produced. However, for qualitative analysis, the visualizations are produced for all three datasets, along with the comparative results with the state-of-the-art schemes. 






\section{Results} \label{Result_Section}
\begin{figure*}[t]
    \centering
    \centerline{\includegraphics[width=0.98\textwidth]{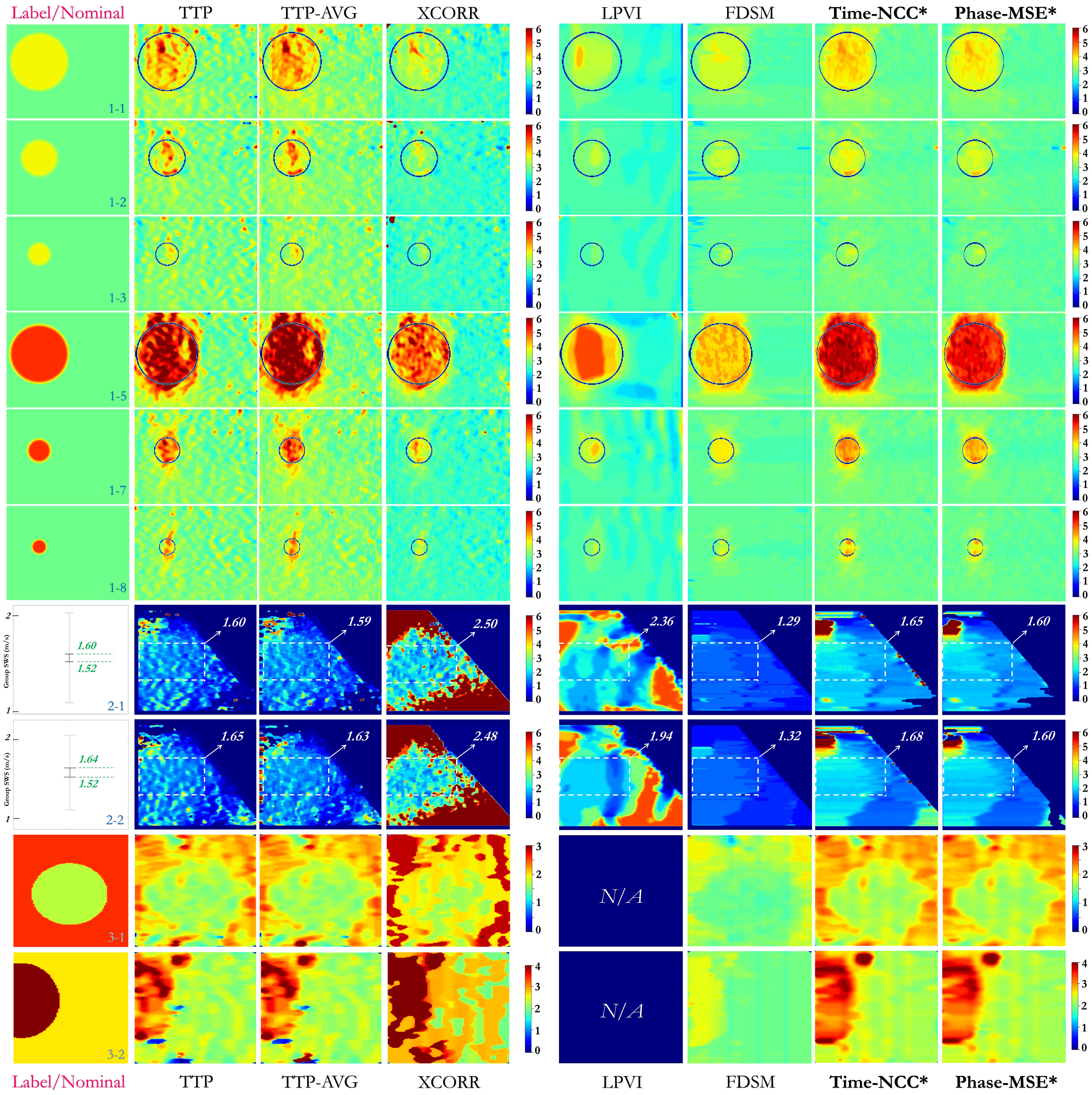}}
    \caption{{Comparative visualization between the constrained optimization-based reconstructions and the state-of-the-art reconstructions on CIRS Model 049 phantoms, RSNA-QIBA-US, and Private phantoms. The two proposed methods are signified with star-marks (*). The color-maps indicate SWS (ranges adjusted as required). The values associated with the dotted-boxes are the mean SWS of the indicated segments.}}
    \label{visu_results1}
\end{figure*}

The performance analysis of the proposed approaches and their comparison with the state-of-the-art techniques are limited to the simulation and CIRS phantom datasets from specialized centers since no clinical data on SWS estimation is yet available publicly. The following are the major observations found from the simulation and experimental outcomes.



\subsection{Simulation Phantom Results} 
The reconstruction results of the simulation phantoms are depicted in figure \ref{homogenous_phantom}. The shown phantoms should have an expected SWS of $2.24ms^{-1}$ and $3.16ms^{-1}$ across the mediums. The estimated mean SWS of a $30.60$~$mm\times17.17$~$mm$ region for both the Time-NCC and Phase-MSE methods are examined. Each technique performs similarly, overestimating the SWS of two elastic mediums by $0.89\%$ and $0.32\%$, respectively (stiffness by $1.79\%$ and $0.64\%$, respectively). The methods performed equivalently in the absence of real-life noise for the simulation data. Both errors are very low for practical usability. Also, the reconstruction outcomes are decently clean and relatively homogeneous. In comparison, the state-of-the-art methods showcase inhomogeneous estimation (figure \ref{homogenous_phantom}: LPVI, FDSM) and overestimation (figure \ref{homogenous_phantom}: XCorr). Although the TTP and TTP-Avg methods output reconstructions with decent mean-SWS, their reconstructions show underestimations (figure \ref{homogenous_phantom}: TTP, TTP-Avg; top-left regions).

\subsection{Experimental Results} \label{Quantitative_Results}
Using Data-1 and 3, a quantitative comparison is provided in table \ref{table:quantitative} based on the aforementioned performance metrics: (i) Mean and Standard-Deviation of the estimated 2D SWS of inclusion ($\hat{\mu}_I, \hat{\sigma}_I$) and background ($\hat{\mu}_B, \hat{\sigma}_B$) regions, (ii) Contrast-to-Noise Ratio ($\mathrm{CNR}$), and (iii) Peak-Signal-to-Noise-Ratio ($\mathrm{PSNR}$). Maintaining the chosen parameters described in Section \ref{exp_setup}, the reconstruction outcomes are evaluated of the proposed scheme as well along with the state-of-the-art techniques: TTP \cite{sandrin2003transient, palmeri2008quantifying, fierbinteanu2009acoustic, chen2013assessment}, TTP-AVG \cite{ahmed2016shear}, Cross-Correlation (XCORR) \cite{tanter2008quantitative}, LPVI \cite{kijanka2018local} and Fourier Domain Shift Matching (FDSM) \cite{rosen2018fourier}. As the diameter of the circular inclusions decreases for the first half of Data-1 ($1-1 \rightarrow 1-4$), the $\mathrm{CNR}$ decreases for all the methods. This is because the inclusion and background stiffness are adjacent to each other. In the second half of Data-1 ($1-5 \rightarrow 1-8$), however, the opposite trend is seen as the stiffness difference between the inclusion and background is higher. Additionally, the inclusion estimations ($\hat{\mu}_I$) degrade significantly for each technique as the inclusion diameters decrease.

Point-wise techniques (TTP, TTP-Avg, XCorr) perform marginally better in estimating the lower diameter inclusions ($1-3,4,7,8$) compared to the techniques which take local coupling into account (LPVI, FDSM, Time-NCC, Phase-MSE). This is due to the boundary of the inclusion being averaged to some extent for the later methods. However, TTP, TTP-Avg, XCorr techniques showcase higher standard deviations in not only inclusion SWS estimations ($\hat{\sigma}_I$) but also background SWS estimations ($\hat{\sigma}_B$).  Among the local coupling techniques, the proposed Time-NCC and Phase-MSE methods produce outputs with the least standard deviation and as a result, these render superior $\mathrm{CNR}$ values. Moreover, the higher $\mathrm{PSNR}$ results for Time-NCC and Phase-MSE indicate low reconstruction error in the normalized sense compared to the other techniques. 

For the Data-3 samples, the shear wave speeds for all regions are underestimated to some degree due to the data being very noisy (illustrated in Section \ref{Discussion_Section}). Nevertheless, the proposed two schemes outperform the state-of-the-art based on $\mu$, $\sigma$, $\mathrm{PSNR}$, and $\mathrm{CNR}$ as a whole. The FDSM technique possesses the lowest $\hat{\sigma}_I$ and $\hat{\sigma}_B$, but underestimates both regions the most as well.  The XCorr method provides a closer estimate of the true value; however, it has very poor $\mathrm{CNR}$ and $\mathrm{PSNR}$. The absence of LPVI's results is due to region merging issue from window selection (see the following section for a detailed explanation).

In figure \ref{visu_results1}, instances of 2D SWS (i.e., $\Upsilon(x,z)$) estimations are visually depicted from each dataset for qualitative analysis and comparison. Some of the Data-1 samples were chosen to show reconstruction results based on stiffness differences and smaller inclusions. The $\Upsilon(x,z)$ of the samples $1-1,2,5,7$ from NCC and MSE provide clean outputs and distinct inclusion-background boundaries. Output boundaries from the samples $1-3,8$ are relatively less clear because SW transitions are low in small diameter inclusions. This explains the relative underestimation of the mean inclusion SWS, $\hat{\mu}_I$ (Time-NCC: $3.27$ and $3.95 ms^{-1}$, Phase-MSE: $3.18$ and $3.83 ms^{-1}$) for the samples $1-3,8$. 
Nevertheless, the Time-NCC and Phase-MSE reconstructions from Data-1 are cleaner than the state-of-the-art techniques.

For Data-2, the nominal SWS of the chosen samples ($2-1,2$) are shown in figure \ref{visu_results1}. The segmented box shown in this figure is the lateral region nearest to the ARF center. Shear waves axially on top and bottom of that particular region are very turbulent. Reconstructions from TTP, TTP-Avg, XCorr, and LPVI are found to be very noisy, especially outside the segmented region. FDSM, Time-NCC, and Phase-MSE outputs are very much cleaner. However, FDSM provides suboptimal SWS mapping inside the segmented region, whereas Time-NCC and Phase-MSE provide SWS mapping close to the nominal ranges.

\begin{figure}[h]
    \centering
    \centerline{\includegraphics[width=0.49\textwidth]{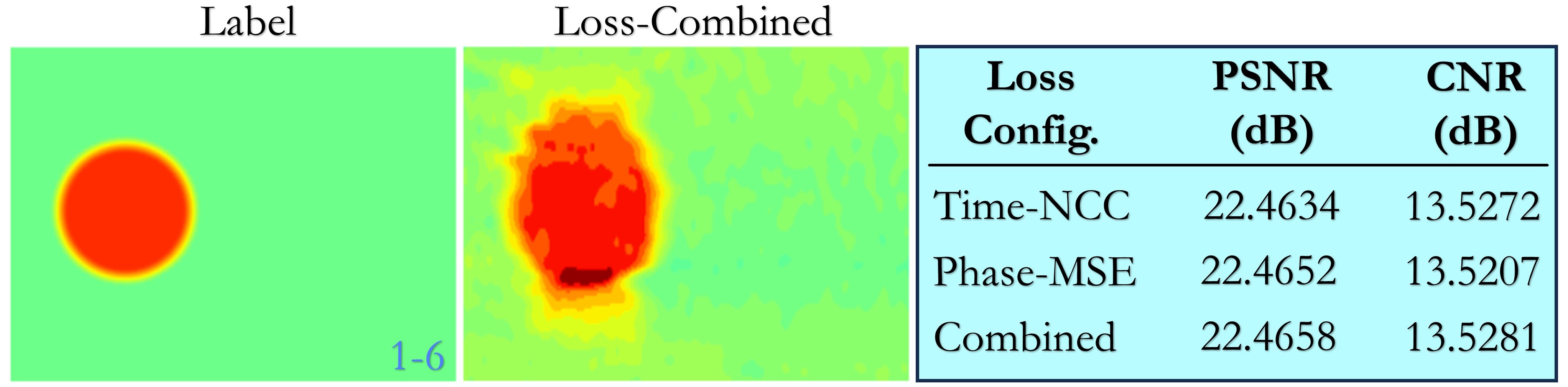}}
    \caption{Reconstruction of sample $1-6$ using Loss-Combined constrained optimization and quality comparison with single loss implementation. }
    \label{combined_loss_result}
\end{figure}

Data-3 was found to be very challenging. Although it has been pre-processed by the data source, it has unexpected signal shapes (explained further in the next section). The corresponding estimations from TTP, TTP-Avg, and XCorr are very noisy (consistent with their low $\mathrm{CNR}$). FDSM, Time-NCC, and Phase-MSE-based outputs are smoother. Additionally, the proposed two methods demonstrate a better contrast in the SWS mapping compared to FDSM. The underestimation of FDSM is consistent with each experimental data.

Finally, figure \ref{combined_loss_result} depicts the result of loss-combined constrained optimization on the sample $1-6$ using $\gamma_1=1.0$ and $\gamma_2=0.2$. Visually the single loss optimization result looks almost identical to the loss-combined version. However, when the $\mathrm{PSNR}$ and $\mathrm{CNR}$ values are inspected, it is found that the combined version performs slightly better.

\begin{figure}[h]
    \centering
    \centerline{\includegraphics[width=0.48\textwidth]{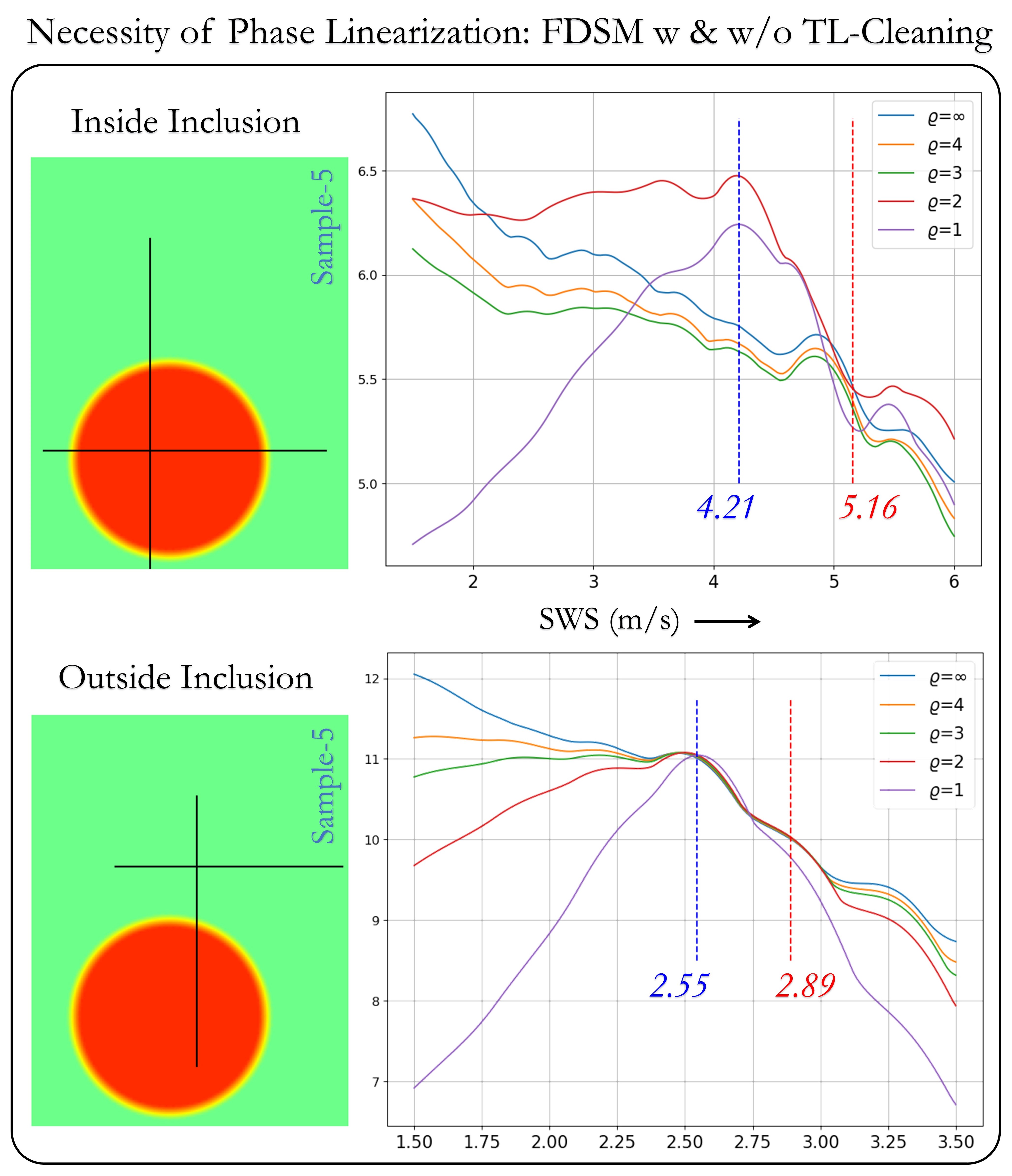}}
    \caption{Necessity of phase-linearization for FDSM optimum SWS estimation. Phase-cleaning is done with ``TL-Cleaning". Increased cleaning improves the peak finding due to phasor-alignment. The vertical blue and red lines indicate the SWS found with $\varrho=1$ cleaning and the true SWS, respectively, for two locations: inside and outside inclusion of Data-1, sample-5 phantom.}
    \label{FDSM_phase_clean}
\end{figure}

\section{Discussion} \label{Discussion_Section}

In this paper, a constrained optimization-based technique has been presented for the shear wave speed estimation in elastic mediums. Section \ref{Result_Section} provides both qualitative and quantitative analysis that the proposed approaches can generate clean and high-quality 2D reconstructions from SWE displacement fields. The Time-NCC and Phase-MSE methods have yielded satisfactory results for both simulation and CIRS phantom experiments. Its effectiveness for $in-vivo$ clinical data is to be explored in future investigations.

It is to be mentioned that to obtain decent results for FDSM, phase linearization was required through ``TL-Plane Cleaning". This provided better $\mathrm{PSNR}$ for FDSM due to its absence of rogue peaks. The reason behind this is that the function defined in (\ref{FDSM_3}) with raw displacement phases does not produce any peaks for maximum argument acquisition. This issue is depicted in Figure  \ref{FDSM_phase_clean}. Without any ``TL-Plane Cleaning" $(\varrho=\infty)$, the phases remain highly nonlinear and no visible peaks from the alignment phasor can be observed.  As the cleaning is increased $(\varrho=4, 3, 2, 1)$, the peaks become more visible. The maximization relies on a higher level of cleaning inside the inclusion compared to the background (i.e., outside inclusion). Even so, the SWS values from the clean versions are underestimated. This provides insight into the fact that linear and clean phase components assist in the optimization process in displacement group alignment or shifting.

\begin{figure}[t]
    \centering
    \centerline{\includegraphics[width=0.47\textwidth]{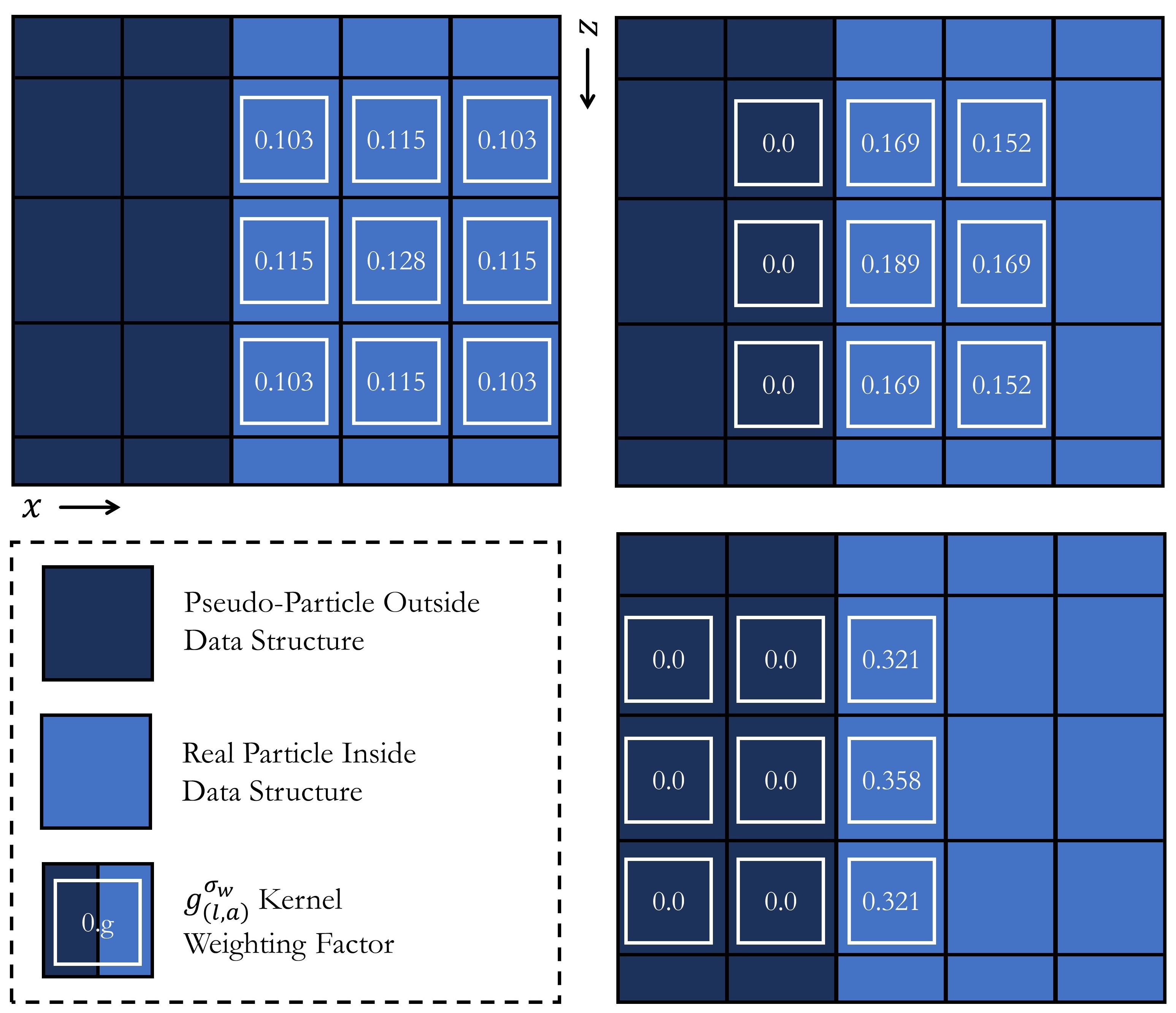}}
    \caption{$g_{l,a}^{\sigma_w}$ kernel modification when capturing an edge of a data structure. For this, a $(3\times3), \sigma_w=0.5$ kernel is demonstrated.}
    \label{kernel_modify}
\end{figure}

Secondly, the results from the LPVI technique are to be addressed. The $\mathrm{CNR}$ of LPVI was calculated to be slightly higher ($10.39dB$) than Time-NCC and Phase-MSE in the case of sample $1-5$ (both $9.06dB$). This is because the estimated background mean-SWS, $\hat{\sigma}_B$, is relatively low in LPVI for using a window shape of near $6.16mm$ which causes a higher degree of smoothing. However, Time-NCC and Phase-MSE-based $\mathrm{PSNR}$ ($18.35$ and $18.70dB$, respectively) are better than that of LPVI ($18.18dB$). As mentioned in section \ref{Intro_section}, LPVI relies heavily on the window size (tiny window: noisy output, large window: small inclusion overlooked). Therefore, smaller inclusions (samples $1-3,7,8$) provide sub-optimal results. For Data-1, the bigger inclusions required a window size near $6mm$ but the smaller ones required below $4.5mm$ with a selected frequency near $F_0=1000Hz$. Furthermore, the results for Data-3 using LPVI have not been generated. Data-3 contains laterally $7mm$ samples as well as $4$ overlapping regions to be merged after reconstruction. Among its $43$ lateral points, the edge-residing ones are significant for merging the regions successfully. Even if LPVI uses a $3mm$ window, a total of $18$ edge pixels will output unreliable results which will not make the region merging possible. For the proposed two methods, the kernel $g_{l,a}^{\sigma_w}$ was modified in such a way so that the edge residing particles still produce reliable results. The modification is shown in figure \ref{kernel_modify}. When the kernel captures an edge, the outside edge residing pixels are zeroed out, and the inside edge weighting factors are normalized to maintain the unity relation from (\ref{gauss}). Kaiser windows were utilized to merge the four reconstructed regions before median filter-based post-processing was introduced. 

\begin{figure*}[h]
    \centering
    \centerline{\includegraphics[width=0.95\textwidth]{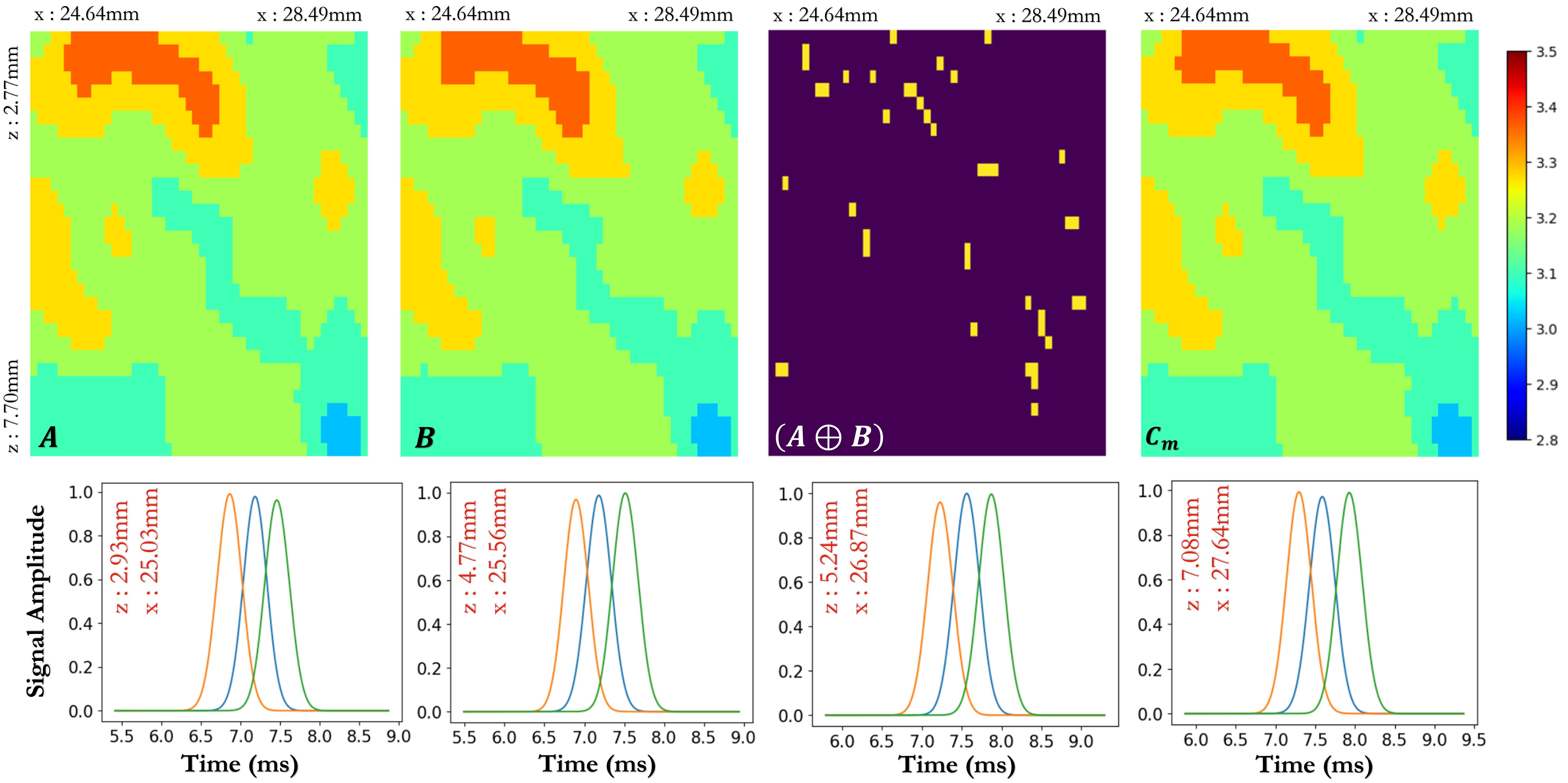}}
    \caption{Outputs form a homogenous portion of the sample $1-5$. $A$: Time-NCC output, $B$: Phase-MSE output, $(A \bigoplus B)$: estimations of SWS which vary between NCC and MSE, $C_m$: Loss-Combined output. The $4$ signal groups are $u_0(n)$, $u_1(n)$ and $u_2(n)$ displacements from $4$ particles instances from $(A \bigoplus B)$.}
    \label{Time_vs_Phase_XOR}
\end{figure*}

In order to emphasize the advantage of the Phase-MSE technique, a homogeneous portion from sample $1-5$ is depicted in figure \ref{Time_vs_Phase_XOR}. Both the Time-NCC (see $A$) output and the Phase-MSE (see $B$) output have been generated using the same setting, yet they are not identical. Their SWS estimations vary in multiple zones (see $A \bigoplus B$). The Phase-MSE background estimation for sample $1-5$ is better than Time-NCC, as seen from table \ref{table:quantitative} (Phase-MSE: 3.22$\pm$0.71, Time-NCC: 3.31$\pm$0.75, Label: 2.89~$ms^{-1}$). This is due to signal groups $u_0(n)$, $u_1(n)$ and $u_2(n)$ not having the same amplitude even after TL-Plane Cleaning. Instances of such signal groups within $A \bigoplus B$ are shown in figure \ref{Time_vs_Phase_XOR}. The unequal amplitudes affect the Time-NCC. However, since only phase components are considered with frequency selectivity ($F_{sig}=\frac{\omega_{sig}}{2\pi}$, from (\ref{reg_phase_01}) and (\ref{reg_phase_20})) in the Phase-MSE technique, the amplitude inequality between displacement groups does not have any impact and the best shift is determined.  

\begin{figure}[h]
    \centering
    \centerline{\includegraphics[width=0.45\textwidth]{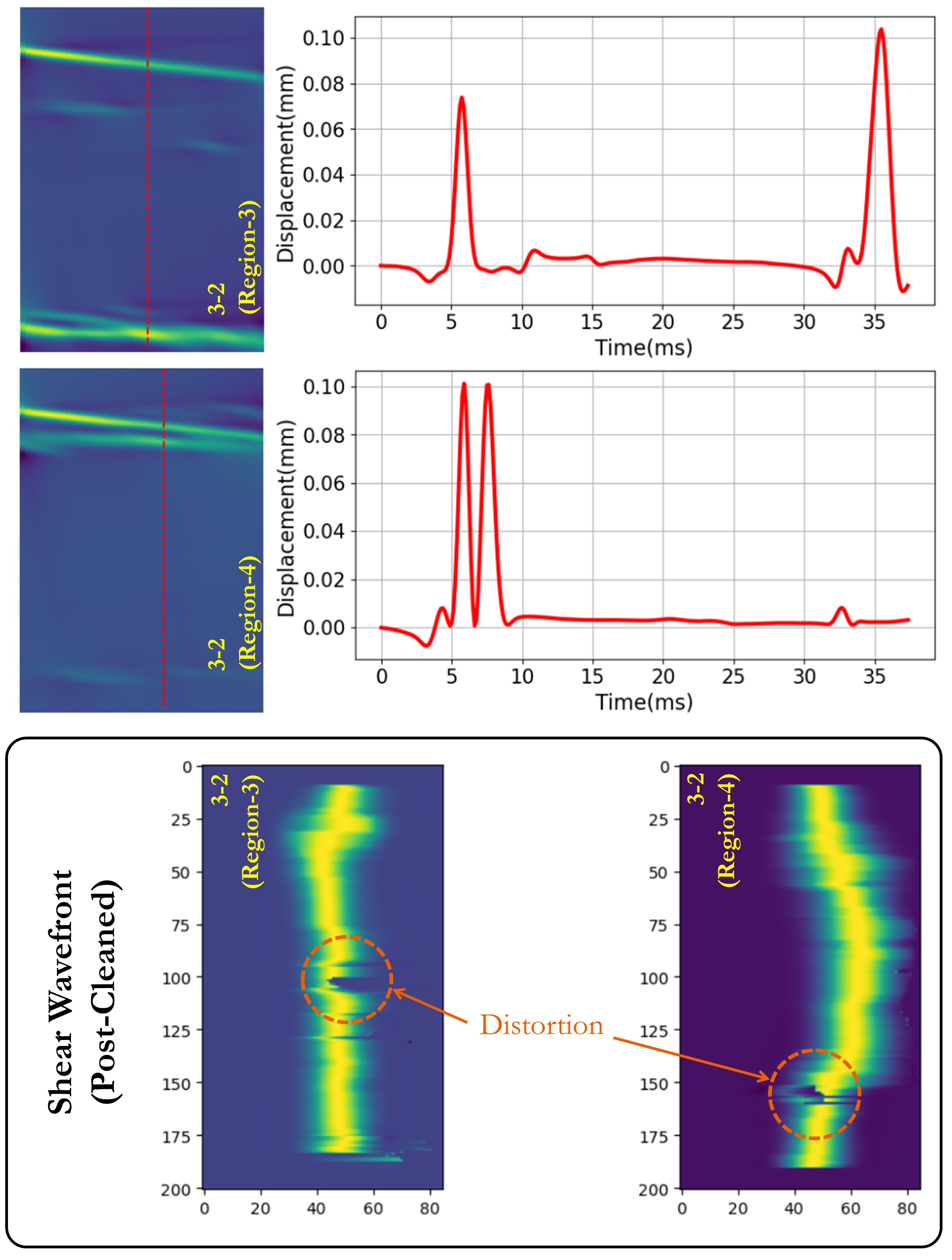}}
    \caption{Some highly noisy displacement fields found in Data-3.}
    \label{private_data_noisy}
\end{figure} 

Another issue to be addressed is that $\mathrm{CNR}$ for the $1-4$ sample is negative for all the methods. Being of very low diameter as well as closer stiffness of inclusion to the background, the change in shear wavefront is very small. For this, the reconstruction has a low difference between the inclusion and background SWS compared to the standard deviation. When results for sample $1-8$ are produced having the same diameter but higher $\hat{\mu}_I$, the $\mathrm{CNR}$ improves due to a higher degree of shear wavefront transition. Nevertheless, the $\mathrm{PSNR}$ and $\mathrm{CNR}$ for the proposed methods outperform the remainder techniques during estimations containing small inclusions.

In particular, the displacement fields of Data-3 are very challenging. This is due to the displacement fields of this particular data being very noisy. Some particles in the data illustrate misleading and unexpected displacement signals, as shown in figure \ref{private_data_noisy}. In such conditions, the regressions performed in the TL-plane cleaning produce errors. For cases similar to figure \ref{private_data_noisy}-top row (a bigger false-peak at the signal ending), the conditioning step of TL-cleaning will perform false normalization. This will adversely influence the next steps. For cases similar to figure \ref{private_data_noisy}-middle row (dual peaks side-by-side), the piece-wise lines fitted in step 3 of TL-plane cleaning will isolate the middle part of the two peaks. This will output a very low amplitude signal. Such signal shapes result in distorted cleaning outcomes, as seen in figure \ref{private_data_noisy}-bottom row and make the estimations sub-optimal.

The limitation of the proposed constrained-optimized methods is the computational expense. For the methods, the 2D reconstruction requires the optimized estimation of each particle in $\Upsilon(x,z)$. A kernel size of higher than $(5\times5)$ will provide clean outcomes but will take an extensive amount of time. The entire pipeline can be coded in C++ to accelerate the production of estimation maps.

\section{Conclusion}

This paper has dealt with noise-robust constrained optimization schemes for estimating Shear Wave Speed (SWS) from shear wave displacement data. Loss functions have been used in both time- and frequency-domains to estimate the optimum arrival time of particles from ARF-induced shear waves. SWS estimation is optimized by minimizing an objective loss function parameterized with a signal group's shift or alignment, subject to equal shift or alignment for neighboring particles with the same stiffness as the center one of the signal group.  In addition to the inherent noise resilience resulting from these constraints, a pre-denosing scheme, called time-lateral plane cleaning, is also implemented to eliminate residual signal oscillations and linearize phase components for better SWS estimation. Simulation and experimental results show superior shear wave image reconstructions in terms of higher CNR and PSNR metrics compared to existing methods. The visual results depict clean backgrounds and distinct inclusion-background boundaries, which are seldom met by existing methods. However, the effectiveness in tissue-mimicking experimental data does not fully account for noise conditions during in-vivo SWE in breast tissue or liver, for example. Nevertheless, the proposed constrained optimization technique shows promise for future in-vivo tissue characterization performance investigation in a clinical setup.

\section{Acknowledgements}
The authors would like to thank BUET for providing Basic Research Grant [grant number 1111202106025] to enhance research at BUET.

\printbibliography

\end{document}